\documentclass[prx,twocolumn,groupedaddress]{revtex4-1}
\usepackage{graphicx,color}
\usepackage{verbatim}
\usepackage{amssymb}   
\usepackage{amsmath}
\usepackage{amsfonts}
\usepackage{mathdots}
\usepackage{hyperref}
\usepackage{footnote}
\usepackage{epsfig}
\usepackage{braket}
\usepackage{bm}
\usepackage{natbib}

\usepackage{multirow}
\usepackage[misc]{ifsym}
\usepackage[hang]{footmisc}
\usepackage{threeparttable}

\begin{document}
	\title{ Kramers Nodal Line Metals}
	\author{Ying-Ming Xie$^1$}
	\thanks{These authors contributed equally to this work}
	\author{Xue-Jian Gao$^1$}
	\thanks{These authors contributed equally to this work}
	\author{Xiao Yan Xu$^2$}
	\author{Cheng-Ping Zhang$^1$}
	\author{Jin-Xin Hu$^1$} 
	\author{Jason Z. Gao$^1$}
	\author{K. T. Law$^1$} \thanks{phlaw@ust.hk}
	\affiliation{$^1$Department of Physics, Hong Kong University of Science and Technology, Clear Water Bay, Hong Kong, China} 	
	\affiliation{$^2$Department of Physics, University of California at San Diego, La Jolla, California 92093, USA}
	\date{\today}
	\renewcommand{\thetable}{\arabic{table}}
	\begin{abstract}
		Recently, it was pointed out that all chiral crystals with spin-orbit coupling (SOC) can be Kramers Weyl semimetals (KWSs) which possess Weyl points pinned at time-reversal invariant momenta. In this work, we show that all achiral non-centrosymmetric materials with SOC can be a new class of topological materials, which we term Kramers nodal line metals (KNLMs). In KNLMs, there are doubly degenerate lines, which we call Kramers nodal lines (KNLs), connecting time-reversal invariant momenta. The KNLs create two types of Fermi surfaces, namely, the spindle torus type and the octdong type.  Interestingly, all the electrons on octdong Fermi surfaces are described by two-dimensional massless Dirac Hamiltonians. These materials support quantized optical conductance in thin films. We further show that KNLMs can be regarded as parent states of KWSs.  Therefore, we conclude that all non-centrosymmetric metals with SOC are topological, as they can be either KWSs or KNLMs.
	\end{abstract}
	
	\maketitle
	\section{\bf{Introduction}}
	The discovery of topological insulators \cite{Kane2,Kane,Bernevig2, Fu,haijun,Qi2,SchnyderA} which possess bulk insulating gap and massless Dirac surface states have inspired intense theoretical and experimental studies in the symmetry and topological properties of electronic band structures. In recent years, a large number of topological insulators and topological semimetals, such as topological crystalline insulators \cite{Hsieh}, higher-order topological insulators \cite{Benalcazar,Schindler,Schindler2,Wang3,Choi}, Dirac semimetals \cite{Rappe,WangZ1,WangZ2,Borisenko, LiuZK1, LiuZK2,Yang2,Xiong,Kim,Wieder,Vishwanath}, Weyl semimetals \cite{Wan,Burkov2,Dai2,Yang,Halasz,Liu,Hirayama,Weng,Huang,Xus,Lv,Soluyanov,Ruan}, nodal line \cite{Burkov,Weng2,Fang3, Bian, Weng3}, nodal chain \cite{Wuquangsheng}, and multifold chiral \cite{Rao,Sanchez,Schroter,Takane,Li2,Yao,Fang,Bradlyn2,Chang2,Tang} topological semimetals, have been discovered. Moreover, systematic ways to diagnose non-trivial band topology based on topological quantum chemistry  and symmetry-based indicators have been developed and a large number of topological materials have been found \cite{PoHC,Kruthoff,Zhangtiantian,TangF,Vergniory}.
	
\begin{figure}
		\centering
		\includegraphics[width=1\linewidth]{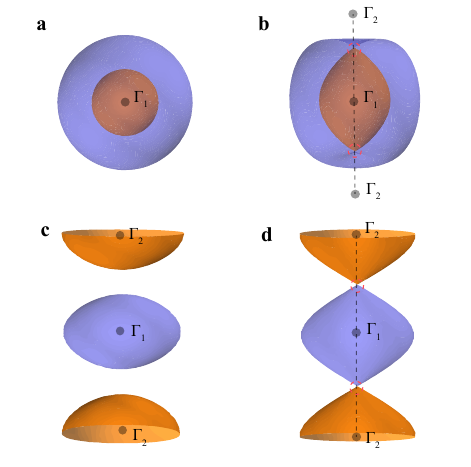}
		\caption{Schematic plot of Fermi surfaces of KWSs and KNLMs. {\bf a} The Fermi surface of a KWS where two Fermi pockets enclose one TRIM. {\bf b} Spindle torus type Fermi surface in a KNLM induced by a KNL (the dashed black line).  {\bf c} The Fermi surface of a KWS where each pocket encloses a different TRIM. {\bf d} Octdong type Fermi surface in KNLMs induced by a KNL.  The gray dots in {\bf a} to {\bf d} indicate the position of TRIMs $\Gamma_1$, $\Gamma_2$. The touching points of the Fermi surfaces are circled by red dashed lines.  }
		\label{fig:fig1}
	\end{figure}

	Recently, the study of   Kramers Weyl semimetals (KWSs) has significantly expanded the family of topological materials \cite{Chang}. It has been stated that in all chiral crystals (crystals which lack mirror or roto-inversion symmetries) with spin-orbit coupling (SOC), each two-fold degenerate time-reversal invariant momentum (TRIM) point is a Weyl point called Kramers Weyl point. Around a Kramers Weyl point, the degeneracy near the TRIM is split along all directions in momentum space by SOC \cite{Samokhin}. Consequently, the Fermi pockets enclosing Kramers Weyl points are split by SOC, and each Fermi pocket possesses nontrivial and opposite Chern numbers, as depicted in Fig.~\ref{fig:fig1}a \cite{Chang}. These KWSs exhibit several novel properties, such as the monopole-like spin texture \cite{Chang,Sakano},  longitudinal magnetoelectric responses \cite{Law, Yoda} and the quantized circular photogalvanic effect \cite{Moore,Chang,Chang2,Flicker,Rees, Ni}.  
	
	In this work, we point out that all non-centrosymmetric achiral crystals (crystals which possess mirror or roto-inversion symmetries ) with SOC possess doubly degenerate lines which connect TRIM points with achiral little group symmetry across the Brillouin zone. The double degeneracy is protected by time-reversal and achiral point group symmetries of the crystal. We call these doubly degenerate lines, Kramers nodal lines (KNLs). It is shown that these KNLs exist in all non-centrosymmetric achiral crystals with SOC. When the Fermi surfaces of materials enclose TRIM points connected by KNLs, we call these materials Kramers nodal lines metals (KNLMs). In Table~\ref{table1}, all the symmorphic space groups (SGs) supporting KNLs are listed, and certain material realizations are identified.

	
	
	Importantly, as long as the Fermi surfaces enclose TRIMs which are connected by KNLs, the KNLs force spin-split Fermi surfaces to touch on the KNLs and create two types of Fermi surfaces, namely, the spindle torus type and the octdong (or hourglass) type as shown in Fig.~\ref{fig:fig1}b and Fig.~\ref{fig:fig1}d, respectively. The band touching points of the Fermi surfaces are described by two-dimensional massless Dirac or higher-order Dirac Hamiltonians \cite{Fang,Yang2,LiuQ,Yu} with the Dirac points pinned at the Fermi energy. In the case of octdong type Fermi surfaces, 
	all the states on the Fermi surfaces are described as two-dimensional massless Dirac fermions. Materials with octdong type Fermi surfaces exhibit linear optical conductivity in the bulk and, in the thin film limit, quantized optical conductivity similar to monolayer graphene due to the massless Dirac fermions \cite{Nair,Li3}.
	
	Furthermore, KNLMs can be regarded as the parent states of KWSs. When the mirror or roto-inversion symmetries are broken, the degeneracies of the KNLs are lifted, and the touching points of the Fermi surface will generally be gapped out and a KNLM becomes a KWS. More specifically, breaking achiral crystal symmetries causes a spindle Fermi surface (Fig.~\ref{fig:fig1}b) to split into two Fermi pockets as shown in Fig.~\ref{fig:fig1}a, and each Fermi pocket carries a net Chern number.  In the case of an octdong Fermi surface (Fig.~\ref{fig:fig1}b), the two Fermi pockets detach from each other and Kramers Weyl points are generated in both pockets, as shown in Fig.~\ref{fig:fig1}c. For illustration, we demonstrate how an isolated Kramers Weyl point near the Fermi energy can be created by breaking the mirror symmetry through strain in BiTeI with a spindle Fermi surface and how this Kramers Weyl point can be detected through the quantized circular photogalvanic effect \cite{Moore}.
	
	From this work, together with the discovery of KWSs, we conclude that all non-centrosymmetric crystals with SOC are topological in nature. They can be either KWSs or KNLMs.

	\begin{table*}
		\begin{threeparttable}[b]
			\caption{ Kramers nodal line metals (KNLMs)  with symmorphic space groups \tnote{*}  	}
			\begin{ruledtabular}\label{table1}
				\begin{tabular}{cccccc}
					Type		&  SG No.         & Point Group 	&              KNLs      &		KW Points        &   Material    \\ 
					\hline
					\multirow{18}{*}{Type I}	& 6,$Pm$     &  $C_{1v}$   &       ($\Gamma$,B,Y,A,Z,C,D,E)\tnote{1}       & -- &      CsIO$_3$         \\
					& 8,$Cm$     &  $C_{1v}$   &       ($\Gamma$,Y,A,M)     & --  &      BiPd$_2$Pb         \\
					& 25,$Pmm2$     &  $C_{2v}$   &  $\Gamma$--Z, Y--T, X--U, S--R   & -- &     CdTe, Bi$_4$Te$_2$Br$_2$O$_9$       \\
					& 38,$Amm2$     &  $C_{2v}$   &       $\Gamma$--Y, T--Z        & -- &    NbS$_2$     \\
					& 42, $Fmm2$     &  $C_{2v}$   &       $\Gamma$--Z, Y--T        & -- &      --         \\
					& 99, $P4mm$     &  $C_{4v}$   &    $\Gamma$--Z, X--R,  A--M    & -- &  PbCsCl$_3$    \\
					& 107, $I4mm$    &  $C_{4v}$   &          $\Gamma$--M ,X--X, (N)           & -- &      In$_2$Te$_3$        \\
					& 115, $P\bar{4}m2$ &  $D_{2d}$   &       $\Gamma$--Z, M--A, X--R       & --  &   PbF$_2$O    \\
					& 156, $P3m1$    &  $C_{3v}$   &          $\Gamma$--A, (M,L)          & -- &     BiTeI      \\
					& 157, $P31m$    &  $C_{3v}$   &          $\Gamma$--A, (M,L)          & -- &   Bi$_2$Pt    \\
					& 160, $R3m$     &  $C_{3v}$   &          $\Gamma$--T, (L,FA)         & -- &      Bi$_2$Te$_3$        \\
					& 174, $P\overline{6}$		&	$C_{3h}$	&	$\Gamma$--A, (M,L) 	& -- &	--\\
					& 183, $P6mm$    &  $C_{6v}$   &        $\Gamma$--A, M--L        & -- &      AuCN         \\
					& 187, $P\bar{6}m2$ &  $D_{3h}$   & $\Gamma$--M, A--L, $\Gamma$--A  & -- &    GeI$_2$, TaN      \\
					& 189, $P\bar{6}2m$ &  $D_{3h}$   & $\Gamma$--K--M, A--H--L, $\Gamma$--A & -- &    Sn$_5$(BIr$_3$)$_2$     \\
					& 215, $P\bar{4}3m$ &    $T_d$    & $\Gamma$--X, $\Gamma$--R, R--M & -- &      Cu$_3$TaTe$_4$         \\
					& 216, $F\bar{4}3m$ &    $T_d$    &          $\Gamma$--L, $\Gamma$--X           & -- &     HgSe, HgTe       \\
					& 217, $I\bar{4}3m$ &    $T_d$    &          $\Gamma$--H          & -- &     TaTl$_3$Se$_4$   \\
					\hline
					\multirow{7}{*}{Type II}		& 35,$Cmm2$     &  $C_{2v}$   &       $\Gamma$--Z, Y--T    &  S, R  &      MnCs$_2$V$_2$Br$_2$O$_6$         \\
					& 44, $Imm2$     &  $C_{2v}$   &          $\Gamma$--X, (S,R)          & T &   AgNO$_2$      \\
					& 81, $P\bar{4}$   &    $S_4$    &          $\Gamma$--Z, M--A         & X, R &   GeSe$_2$      \\
					& 82, $I\bar{4}$   &    $S_4$    &          $\Gamma$--M           & N, X &   CdGa$_2$Te$_4$, Cr$_2$AgBiO$_8$  \\
					& 111, $P\bar{4}2m$ &  $D_{2d}$   &       $\Gamma$--Z, M--A        & X, R & Ag$_2$HgI$_4$ \\
					& 119, $I\bar{4}m2$ &  $D_{2d}$   &          $\Gamma$--M,  (N)        & X  &      TlAgTe$_2$          \\
					& 121, $I\bar{4}2m$ &  $D_{2d}$   &          $\Gamma$--M,  X--X         & N &     Cu$_3$SbS$_4$      
					
					
				\end{tabular}
				
			\end{ruledtabular}
			\begin{tablenotes}[flushleft]
				\item [*] Here we enumerate  symmetry allowed KNLs in symmorphic space groups. The definitions of TRIMs follow the conventions given in Bilbao Crystallographic Server \cite{Elcoro}. Some of the representative materials hosting KNLs are identified with the assistance of the Materials Project \cite{Jain2} and the Topological Material Database \cite{Vergniory}.
				\item [1] The TRIMs in the parentheses are connected by the KNLs which are not along the high symmetry lines, such as ($\Gamma$,A), (Y,M) in SG No. 8 (\textit{Pm}) and (M,L) in SG No. 156 (\textit{P3m1}). 
			\end{tablenotes}
		\end{threeparttable}
	\end{table*}
	\section{\bf Results}

	\subsection{ Emergence of Kramers nodal lines from TRIMs with achiral little group symmetry}
	In this section, we demonstrate how nodal lines emerge out of a TRIM with achiral little group symmetry  (which contains mirror or roto-inversion). According to Kramers theorem, each electronic band is  at least doubly degenerate at a TRIM $\mathbf{k}_0$, where $\mathbf{k}_0=-\mathbf{k}_0+\mathbf{G}_i$, and $\mathbf{G}_i$ denotes a reciprocal lattice vector. We first focus on the cases that the energy bands are two-fold degenerate at TRIM points, and the cases with four-fold degeneracy are discussed in the Method Section.  In general, the energy bands near the TRIM $\mathbf{k}_0$ with little group symmetry (Supplementary Note 2) $\mathcal{G}_{\mathbf{k}_0}$ can be described by a Hamiltonian 
	\begin{equation}
	H(\mathbf{k})=f_0(\mathbf{k})+ \bm{f}(\mathbf{k})\cdot\mathbf{\sigma},
	\label{eq:1}
	\end{equation}
	where $\mathbf{k}$ is measured from the TRIM $\mathbf{k}_0$, $\mathbf{\sigma}$ are Pauli matrices operating on the spin space,   $\bm{f}(\mathbf{k})\cdot\mathbf{\sigma}$ denotes the SOC and the eigenvalues of $H(\mathbf{k})$ can be written as $E_{\pm}(\mathbf{k})=f_0(\mathbf{k})\pm |\bm{f}(\mathbf{k})|$.  
	
	As $H(\mathbf{k})$ respects the time-reversal symmetry $\mathcal{T}=i\sigma_y K$ ($K$ is the complex conjugate operation) and the little group  symmetry $\mathcal{G}_{\mathbf{k}_0}$,  $\bm{f}(\mathbf{k})$ satisfies the symmetry constraints
	\begin{equation}
	\bm{f}(\mathbf{k})=-\bm{f}(-\mathbf{k}), \bm{f}(\mathbf{k})=\text{Det}(R)R^{-1}\bm{f}(R\mathbf{k}), \label{Eq2}
	\end{equation}  
	where $R\in \mathcal{G}_{\mathbf{k}_0}$.
	
	For illustration, we analyze the case where $\bm{f}(\mathbf{k})$ is linear in $\mathbf{k}$,  \textit{i.e.}, $\bm{f}(\mathbf{k})=\hat{M}\mathbf{k}$, where $\hat{M}$ is a matrix. A more general proof is provided in the Supplementary Note 2. According to Eq.~(\ref{Eq2}), $\hat{M}$ satisfies $\hat{M}=\text{Det} (R) R^{-1}\hat{M}R$.  Denoting $\mathbf{n}_j$ and $\epsilon_j$ as the eigenstates and the eigenvalues of matrices $\hat{M}$ satisfying  $\hat{M}\mathbf{n}_j=\epsilon_j\mathbf{n}_j$, and decomposing  the momentum  $\mathbf{k}$  with the new basis  as $\mathbf{k}=\sum_jp_{j}\mathbf{n}_j$, one finds
	\begin{equation}
	\bm{f}(\mathbf{k})=\sum_jp_j\epsilon_j\mathbf{n}_j.
	\end{equation}
	In general, for a TRIM with a little group symmetry which is chiral, $\text{Det}(\hat{M})\neq 0$, namely $\epsilon_j$ are all finite.  In this case,  $|\bm{f}(\mathbf{k})|>0$ as long as $\mathbf{k}$ is not at the TRIM, which results in a fully split Fermi surface as shown in Fig.~\ref{fig:fig1}a and makes the TRIM a Kramers Weyl point as pointed out in Ref.~\cite{Chang}.  In contrast, for a TRIM with an achiral little group,  there exists at least one  mirror or roto-inversion operation  $\tilde{R}$ with $\text{Det}(\tilde{R})=-1$ such that $\text{Det}(\hat{M})=0$, implying that at least one of $\epsilon_j$ is zero. Without loss of generality, taking $\epsilon_3=0$,  one obtains
	\begin{equation}
	\bm{f}(\mathbf{k})=p_1\epsilon_1\mathbf{n_1}+p_2\epsilon_2\mathbf{n_2}
	\label{eq:3}.
	\end{equation}
	$\bm{f}(\mathbf{k})$ vanishes when the momentum $\mathbf{k}$ is fixed to be along the direction of null vector $\mathbf{n_3}$ where $p_1=p_2=0$ and $\mathbf{k}=p_3\mathbf{n_3}$. In this case,  $E_{+}(\mathbf{k})$ and $E_{-}(\mathbf{k})$ are degenerate along the $\mathbf{n_3}$-direction. The line $\mathbf{k}=p_3\mathbf{n_3}$ is an example of a degenerate line coming out of TRIMs. The degeneracy is protected by time-reversal symmetry and the achiral little group symmetry. We called these lines, KNLs. It is important to note that KNLs create touching points on the Fermi surface at any Fermi energy as long as the Fermi surface enclose TRIMs with achiral little groups, as depicted schematically in Fig.~\ref{fig:fig1}.  Interestingly, these touching points, which are always pinned at the Fermi energy, are two-dimensional Dirac points or higher-order Dirac points \cite{Fang,Yang2,LiuQ,Yu} with non-trivial topological properties (Supplementary Note 3). The general form of the $\mathbf{k\cdot p}$ Hamiltonians of all non-centrosymmetric achiral point groups and the directions of KNLs emerging out of the TRIM are summarized in the Method Section. Beyond the $\mathbf{k}\cdot \mathbf{p}$ analysis, we showed in the Supplementary Note 2 that for a general $\bm{f}(\mathbf{k})$, the KNLs are guaranteed to  lie within the mirror planes or along the roto-inversion axis of $S_3$, $S_4$ symmetry. It is further shown that  a KNL emerging from one TRIM has to connect with another TRIM with an achiral little group (Supplementary Note 2).
	
	
	
	
	
	\begin{figure*}
		\centering
		\includegraphics[width=1\linewidth]{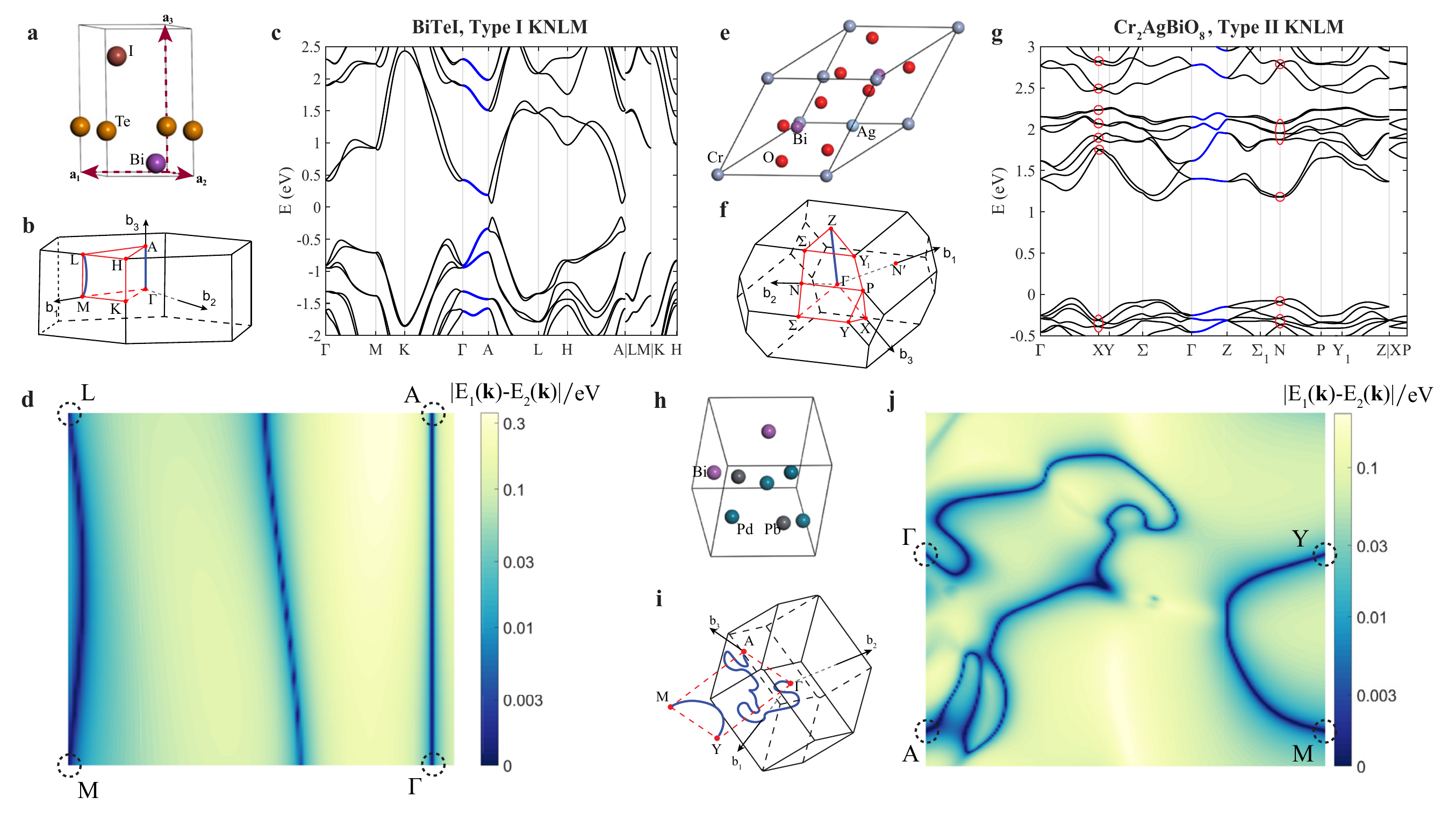}
		\caption{Representative materials with KNLs. {\bf a} to {\bf j} The crystal structure, the first Brillouin zone, and KNLs of BiTeI (SG No.~156, $P3m1$), Cr$_2$AgBiO$_8$ (SG No.~82, $I\bar{4}$) and BiPd$_2$Pb (SG No.~8, $Cm$). {\bf c} and {\bf g} are the band structures of BiTeI and Cr$_2$AgBiO$_8$, respectively, where the KNLs are highlighted as blue lines, and the crossing points within the red circles of {\bf f} are KW points. These KNLs are also marked out by solid blue lines in the 3D first Brillouin zone. {\bf d} and {\bf j} show the DFT-calculated energy difference of two selected SOC-split bands $|E_1(\mathbf{k})-E_2(\mathbf{k})|$ (in units of eV) on a mirror-invariant $k$ plane for BiTeI and BiPd$_2$Pb,  respectively. The dark green lines that connect two TRIMs (dashed circles) are KNLs on this mirror plane.}
		\label{fig:fig2}
	\end{figure*}

	\subsection{ Kramers nodal lines in achiral crystals}
	
	In the previous section, we demonstrated how KNLs emerge out of TRIMs. In this section, we study how KNLs connect different TRIMs in non-centrosymmetric achiral crystals. While most KNLs connect TRIMs along high symmetry lines, some KNLs connect TRIMs through general points in the mirror plane (such as for TRIMs with $C_{1v}$ little groups).
	
	To identify the KNLs joining TRIMs along high symmetry lines, we make use of the compatibility relations of double-valued space groups \cite{Bradley,Elcoro}, which are defined by
	\begin{equation}
	\chi(D^{(\Gamma_1)}_{\mathcal{G}_1}(R))=\sum_j \chi(D^{(\Gamma_j)}_{\mathcal{G}_2}(R))\label{Eq_compatibility},
	\end{equation}
	where  $\chi$ is the character of a symmetry operation $R$ in a specific representation, $\mathcal{G}_1$ and $\mathcal{G}_2$ are the little groups of the TRIM and a high symmetry line respectively and $D^{(\Gamma_j)}_{\mathcal{G}_i}(R)$ is the $j$th irreducible representation of the symmetry operation $R\in \mathcal{G}_i$. For example, for the well-studied 3D Rashba material BiTeI (SG No.~156, $P3m1$), the little groups of the TRIM $\Gamma$, A and the high symmetry line $\Delta$ connecting these two TRIMs are all $C_{3v}$. By identifying the irreducible representations of the unitary symmetry operations $m_{010}$ and $C_3$ at $\Gamma$, A and $\Delta$ (see Supplementary Note 4 for details), we show that the two-dimensional double-valued irreducible representations $\overline{\Gamma}_6$--$\overline{\Delta}_6$--$\overline{\text{A}}_6$ are compatible. This explains all the KNLs $\Gamma$--A observed in the band structure of BiTeI shown in Fig.~\ref{fig:fig2}c (labeled with blue color). This result is also consistent with the $\mathbf{k\cdot p}$ Hamiltonian analysis that a KNL emerges out of the $\Gamma$ point along the $z$-direction (see the Method Section).
	
	Based on the compatibility relations, we identified all the KNLs which are along the high symmetry lines in non-centrosymmetric crystals with symmorphic space groups. The results are summarized in Table~\ref{table1}.  We found non-centrosymmetric achiral crystals with point groups $C_{2v}$, $S_4$, $C_{4v}$, $D_{2d}$, $C_{3v}$, $C_{3h}$, $C_{6v}$, $D_{3h}$, $T_d$ support KNLs along high symmetry directions.  These lines are contained within the mirror plane or along the roto-inversion axis.  Some representative materials with  KNLs are listed in Table~\ref{table1}. For example, for space group 216, there are KNLs along the high symmetry lines between $\Gamma$ and L points as well as between $\Gamma$ and X points. These KNLs are labeled as $\Gamma$--L and $\Gamma$--X, respectively, in Table~\ref{table1}. Materials with this property include semimetals HgTe and HgSe. For further illustration, the band structures of BiTeI (SG No.~156, $P3m1$) and Cr$_2$AgBiO$_8$ (SG No.~82, $I\bar{4}$) are shown in Fig.~\ref{fig:fig2}.  Evidently, there are KNLs (labeled with blue color) along the high symmetry lines.

	Although most KNLs reside on high symmetry lines, there are exceptions if the little group of the TRIM is $C_{1v}$. As shown in the previous section, $C_{1v}$ is achiral so that there must be KNLs emerging from TRIMs.  For example, the little groups of TRIMs M and L in BiTeI are the achiral $C_{1v}$, yet there are no KNLs along high symmetry lines coming out from M or L, as shown in Table~\ref{table1}. However, by carefully checking the energy bands on the whole mirror plane,  as shown in Fig.~\ref{fig:fig2}d (and schematically shown in Fig.~\ref{fig:fig2}b), we indeed found a KNL that connects M, L  within the mirror plane which is denoted as (M,L) in Table~\ref{table1}. Therefore, all TRIMs in BiTeI are connected by KNLs as expected.
	
	On the other hand,  there exist TRIMs with chiral little group symmetry, such as the X and N points in achiral KNLM Cr$_2$AgBiO$_8$. Therefore, the Bloch states for each band near X and N points in Cr$_2$AgBiO$_8$ are described by Kramers Weyl fermions, as highlighted in Fig.~\ref{fig:fig2}g. As demonstrated in the Supplementary Note 7, Fermi arcs originating from these Kramers Weyl points emerge on (001) surfaces of Cr$_2$AgBiO$_8$.   As summarized in Table~\ref{table1}, among the 25 non-centrosymmetric achiral symmorphic space groups, 18 of them are classified as Type I achiral crystals in which all the TRIMs are connected by KNLs. In contrast, the remaining seven space groups further support Kramers Weyl points, and they are classified as Type II achiral crystals.
	\begin{figure*}
		\centering
		\includegraphics[width=1\linewidth]{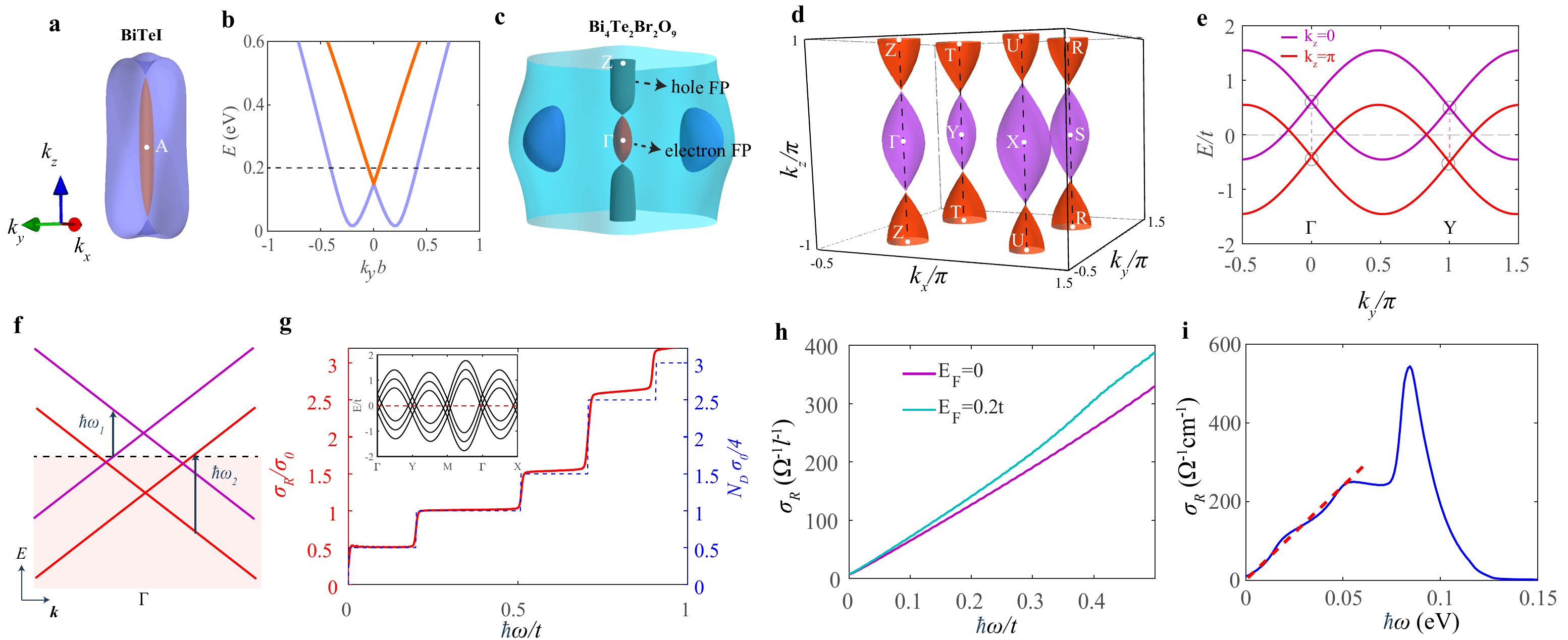}
		\caption{Spindle torus and octdong Fermi surfaces. {\bf a} The Fermi surface of  BiTeI  with Fermi energy $E_F=0.2$ eV, which cuts through the KNL $\Gamma$--A. The inner (orange) and outer (purple) Fermi pockets (FP) together form a spindle torus. The energy dispersion at a fixed $k_z$ indicated by the dashed line is shown in {\bf b}. {\bf b} The Rashba-like energy dispersion for a fixed $k_z$. {\bf c} The Fermi surface  of Bi$_4$Te$_2$Br$_2$O$_9$ (SG No.~25, $Pmm2$) with Fermi energy $E_F=0.05$ eV, which cuts through the KNL $\Gamma$--Z. The labeled hole and the electron Fermi pockets together form an octdong type Fermi surface. {\bf d} The Fermi surface from the two-band tight-binding model $\mathcal{H}_0(\mathbf{k})$ with $m_x=0.05t, m_y=0.05t, m_z=0.5t, v_x=t, v_y=t, E_F=0$ and $t=1$ as the unit of the hybridization energy. The positions of TRIMs depicted are all connected by four KNLs in the $k_z$-direction. {\bf e} The energy dispersion  for a fixed $k_z=0$ (purple) and $k_z=\pi$ (red) in {\bf d}. {\bf f} Schematic plot of optical excitations that contribute to the optical conductivity for the hole-type  (electron-type) Dirac fermions with onset frequency $\omega_1$ ($\omega_2$).  The horizontal dashed line denotes the position of Fermi energy.   {\bf g} The optical conductivity $\sigma_{R}$ (left axis) and estimated  optical conductivity $N_D\sigma_0/4$ (right axis) versus  frequency $\omega$ for a three-layer slab, where the number of Dirac points $N_D=\frac{1}{2}\sum_{\Gamma,n} \theta(\hbar\omega-|2E_{\Gamma,n}|)$ with $\theta$ as the Heaviside step function, $n$ as band index and $\Gamma$ labeling four TRIMs. The inset figure in {\bf g} shows the band structure of this trilayer slab. {\bf h}  The bulk optical conductivity for the model material with octdong Fermi surface at $E_F=0, 0.2t$ with $\eta=0.002t$ and temperature $T=0.01 t$. Here $l^{-1}=\frac{2\pi}{\tilde{a}}$ cm$^{-1}$ with $\tilde{a}=a/$\AA\ and $a$ as the lattice constant.  {\bf i} The bulk optical conductivity  for  Bi$_4$Te$_2$Br$_2$O$_9$ with  $\eta=1$ meV  and temperature $T=10$K.   The slight deviation from linear dependence (red dashed line) for Bi$_4$Te$_2$Br$_2$O$_9$ is due to the presence of the extra trivial pockets (blue pockets in {\bf c}).   }
		\label{fig:fig3}
	\end{figure*}	
	
	One interesting example of KNLs can be found in BiPd$_2$Pb (SG No.8, $Cm$, point group $C_{1v}$), which exhibits large SOC-induced band splitting $\sim$100meV (see Supplementary Note 7  for the band structure). The lattice structure and the Brillouin zone is shown in Fig.~\ref{fig:fig2}h and Fig.~\ref{fig:fig2}i, respectively. In Fig.~\ref{fig:fig2}j, we select two bands which are degenerate on the TRIMs and plot the energy difference with respect to momentum $\mathbf{k}$ in the mirror plane (see the detail band structure in Supplementary Note 7). Remarkably, there are two KNLs,  ($\Gamma$--A) and (Y--M), lying on this mirror plane as expected. The schematic plot of the KNLs on the mirror plane is depicted in Fig.~\ref{fig:fig2}i.  While KNLs along high symmetry lines can easily be found in standard band structure calculations, this kind of irregular KNLs coming out of TRIM with $C_{1v}$ little groups can easily be missed.


	\subsection{ Spindle torus type and octdong type Fermi surfaces}

	In this section, we point out an important physical consequence of the KNLs, namely, KNLs force SOC split Fermi surface to touch. Interestingly, there are two kinds of Fermi surface touchings which can satisfy the doubly degenerate requirement of KNLs. The first type is the spindle torus Fermi surface formed by the touching of two electron Fermi pockets, as illustrated schematically in Fig.~\ref{fig:fig1}b, in which the KNL forces the two SOC split Fermi pockets to touch. The spindle torus Fermi surfaces are rather common in achiral crystals with strong SOC. It is well-known that BiTeI possesses this kind of Fermi surface \cite{Murakawa}, and we explain here that the origin of the Fermi surface touching is indeed enforced by the  $\Gamma$--A KNL, as illustrated in Fig.~\ref{fig:fig3}a. To understand the properties of the electrons on spindle Fermi surfaces, we use BiTeI as an example and note that with a fixed $k_z$, the electrons on the Fermi surface are described by a two-dimensional Rashba Hamiltonian as illustrated in Fig.~\ref{fig:fig3}b \cite{Ishizaka, Bahramy}.  In this work, we point out that almost all non-centrosymmetric achiral crystals with strong SOC have similar properties even though the Fermi surfaces can be more complicated. In the case of hole-doped HgTe and HeSe, for example, three KNLs come out of the $\Gamma$ point and result in six Fermi surface touching points, as illustrated in the Supplementary Note 7.
	
	The second type of Fermi surface touchings which satisfies the degeneracy requirement on the KNLs is the octdong type Fermi surface. In this case, one electron Fermi pocket and one hole Fermi pocket touch along the KNL, as illustrated in Fig.\ref{fig:fig1}b schematically and in Fig.~\ref{fig:fig3}c using the realistic band structures of Bi$_4$Te$_2$Br$_2$O$_9$ (SG No.~25, $Pmm2$, point group $C_{2v}$). In Bi$_4$Te$_2$Br$_2$O$_9$, there is an octdong Fermi surface near the $\Gamma$ point, and the KNL is along the $\Gamma$--Z direction. It is important to note that this Fermi surface touching is not accidental but forced by the KNL. As the chemical potential changes, the relative size of the electron and hole pockets changes and the band touching point moves along the KNL. Importantly, for a fixed $k_z$ along the nodal line direction, the electrons on the octdong Fermi surface are described by two-dimensional massless Dirac fermions on the whole Fermi surface. 

	The octdong Fermi surface as well as the trivial Fermi sheet of Bi$_4$Te$_2$Br$_2$O$_9$ in Fig.~\ref{fig:fig3}c can be captured by  a simple tight-binding Hamiltonian, which satisfies the space group symmetry SG No.~25 ($Pmm2$). The effective Hamiltonian can be written as
	\begin{equation}\label{Eq_model_octdong}
	\mathcal{H}_0(\mathbf{k})=\sum_jm_j\cos(k_j)+v_x\sin k_x\sigma_x+v_y\sin k_y\sigma_y,
	\end{equation}
	where $j=x,y,z$, $\sigma$ are Pauli spin matrices.  As illustrated in Fig.~\ref{fig:fig3}d, it is interesting to note that symmetry allows the crystal to possess pure octdong Fermi surfaces when SOC is further enhanced. Unfortunately, we have yet to identify realistic materials with pure octdong Fermi surfaces.
	
	To understand the novel properties of octdong Fermi surfaces, we first study the optical properties of a system with the octdong Fermi surface only as depicted in Fig.~\ref{fig:fig3}d. The cases with additional trivial Fermi surfaces will be discussed later. We note that in the case of Fig.~\ref{fig:fig3}d, all the electrons on the Fermi surface are described by two-dimensional massless Dirac fermions with Dirac points located on the KNLs. The massless Dirac energy dispersions at $k_z=0$ and $k_z=\pi$ are depicted in Fig.~\ref{fig:fig3}e. It is clear from Fig.~\ref{fig:fig3}e that the energy bands cross at $\Gamma$ and Y points which are Dirac points. Dirac points corresponding to general $k_z$ lie along the dashed lines in Fig.~\ref{fig:fig3}e between the two Dirac points highlighted by circles. In other words, all the states on the octdong Fermi surface can be described by two-dimensional massless Dirac Hamiltonians, and the energy of the Dirac points is determined by $k_z$.  We expect the large number of Dirac electrons on octdong surfaces possess novel physical properties.
	
	To illustrate this, we calculate the optical conductivity $\sigma_{R}(\omega)\equiv \text{Re}(\sigma_{xx}(\omega))$ for a thin film of material with the octdong Fermi surface using a tight-binding version of the effective Hamiltonian (Eq.~(\ref{Eq_model_octdong})). The energy spectrum of such a trilayer thin film is shown in the insert of Fig.~\ref{fig:fig3}g which can be effectively described by multiple massless Dirac Hamiltonians. Applying the Kubo formula, the optical conductivity can be written as
	
	\begin{eqnarray}\label{optical_kubo}
	&&\sigma_{R}(\omega)=\frac{ e^2}{\hbar V}\sum_{\mathbf{k}}\sum_{i\neq j} \frac{f(\epsilon_i(\mathbf{k}))-f(\epsilon_j(\mathbf{k}))}{\epsilon_i(\mathbf{k})-\epsilon_j(\mathbf{k})}\cdot\nonumber\\
	&&|\braket{i,\mathbf{k}|\hat{v}_{x}|j,\mathbf{k}}|^2\text{Im}(\frac{1}{\hbar\omega+i\eta+\epsilon_i(\mathbf{k})-\epsilon_j(\mathbf{k})}),
	\end{eqnarray}
	where $\omega$ is the frequency of the incident light, $V$ is the volume (area) for a bulk (thin film) sample, $i,j$ are the band indices, $f$ is the Fermi-Dirac distribution function, $\eta$ originating from the effect of carrier damping  is assumed to be a constant,   and $\hat{v}_x=\partial \mathcal{H}_0/\partial k_x$ is the velocity operator. As shown in Fig.~\ref{fig:fig3}g, remarkably,  the optical conductivity is quantized and shows plateau structures. The quantization is similar to monolayer graphene which exhibits quantized optical conductivity of $\sigma_0=\pi e^2/2h$  in the frequency range $\omega>2|\mu|$, with $\mu$ being the chemical potential measured from the Dirac point \cite{Kuzmenko,Nair,MacDonald}. To understand the plateau structure, we note that different Dirac points of the thin film have different activation frequencies at which light can excite occupied states into empty states, as depicted in Fig.~\ref{fig:fig3}f. As the optical frequency increases, more and more optically activated Dirac points contribute to quantized optical conductivity and result in the plateau structure. By counting the number of Dirac points $N_D$  within half of the optical frequency $\omega$, we obtain the quantized  plateaus (blue dashed line in Fig.~\ref{fig:fig3}g) that is consistent with the one calculated with the Kubo formula (Eq.~(\ref{optical_kubo})).  This clearly demonstrates the novel properties of materials with octdong Fermi surfaces. The deviation from the quantization values at higher frequencies is due to the deviation from the Dirac energy spectrum at energy far from the Dirac points.
	
	
	\begin{figure*}
		\centering
		\includegraphics[width=1\linewidth]{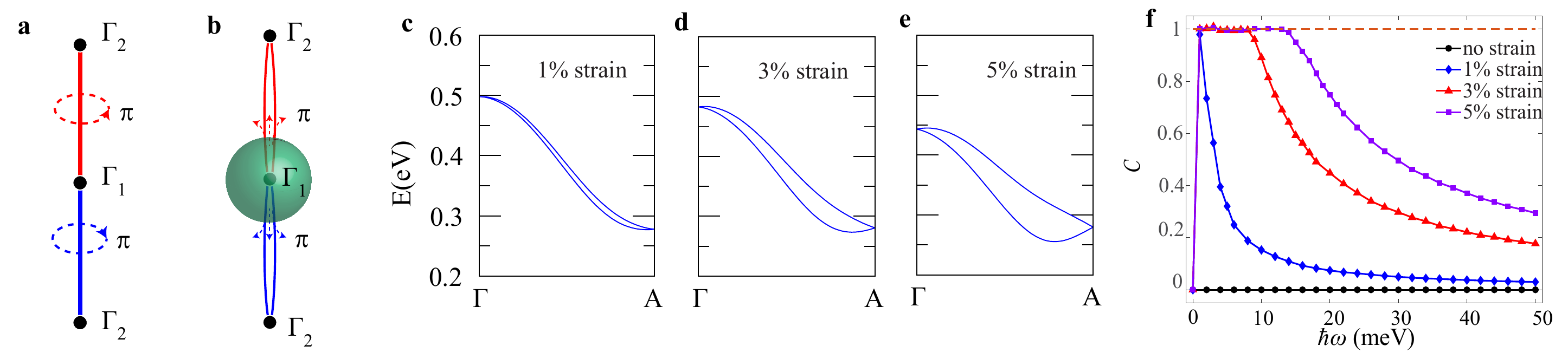}
		
		\caption{Strain-induced Kramers Weyl fermions. {\bf a} Schematic plot of a KNL (solid line) carrying Berry flux $\pi$. {\bf b} The Berry flux emerges from TRIMs when the degeneracy of the KNL is lifted. The total flux through  a sphere (in green) that enclose the TRIM is $2\pi$.  
			{\bf c}, {\bf d} and {\bf e} show the splitting along $\Gamma$--A with $1\%$, $3\%$ and $5\%$ strain strengths, respectively. {\bf f}  The chiral charge $\mathcal{C}$  versus light frequency  $\omega$,   calculated at four different strain strengths: no strain (in black), $1\%$ strain (in blue), $3\%$ strain (in red) and $5\%$ strain (in purple). }
		\label{fig:fig4}
	\end{figure*}
	
	The number of two-dimensional massless Dirac fermions are expected to scale with the system size. In the bulk limit, the optical conductivity with octdong Fermi surfaces is linearly proportional to the optical frequency due to the large number of two-dimensional massless Dirac fermions, as denoted by the linear line in Fig.~\ref{fig:fig3}h.  Importantly, the onset frequency for this linear line is pinned at zero regardless of chemical potential (Fig.~\ref{fig:fig3}h). The underlying reason is that  those  touching points on the octdong Fermi surface always manifest as massless Dirac points right at Fermi energy.  This is substantially different from the linear optical conductivity shown in Weyl \cite{Ashby,Xu}, Dirac semimetals \cite{Tabert,Neubauer,Chen} and multi-fermions \cite{Grushin} where the onset frequency depends on how far the chemical potential is away from the Weyl or Dirac points.  Moreover, as shown in Fig.~\ref{fig:fig3}i, in the case of  the coexistence of an octdong Fermi surface and trivial Fermi surfaces in Bi$_2$Te$_2$Br$_2$O$_9$, the optical conductivity, which is calculated from realistic tight-binding models constructed with Wannier orbitals from DFT calculations (Supplementary Note 7), also shows such linear increase,  although it is limited to a relatively smaller frequency range.   When the optical frequency is high, transitions appear between states which are far from the Dirac points, and the linear behavior of the optical conductivity is lost. To experimentally demonstrate this linear optical conductivity in KNLMs, the incident direction of light should be parallel to the KNLs, and the Drude response that gives a peak near zero frequency  needs to be subtracted \cite{LeeJ}. 
	
	

	\subsection{KNLMs as the parent states of Kramers Weyl materials}
	
	
	In this section, we point out that KNLMs are parent states of KWSs and one can obtain KWSs from KNLMs through lattice symmetry breaking. To understand the relation between KNLMs and KWSs, we note that the KNLs are doubly degenerate lines connecting TRIMs. A plane in the Brillouin zone intercepting a KNL can be described by a two-dimensional massless Dirac Hamiltonian with Berry curvature concentrated at the Dirac point. When a Bloch electron moves around a KNL adiabatically, it acquires a quantized Berry phase of $m\pi$ mod $2\pi$ (Supplementary Note 3), and one can regard a KNL carrying Berry curvature flux of $\pi$ as a Dirac solenoid, as illustrated in Fig.~\ref{fig:fig4}a. It is important to note that the Berry curvature on the opposite sides of a TRIM should have opposite signs  because of time-reversal symmetry such that the Dirac solenoids \cite{Mikitik} manifested by KNLs do not have classical analogues.  When the  symmetries (such as the mirror or the roto-inversion) of a crystal are broken, the degeneracy of the KNLs is lifted, and it is possible to define a nondegenerate Fermi surface enclosing a TRIM. As depicted in Fig.~\ref{fig:fig4}b, the Berry flux coming out of a TRIM is quantized. Therefore, the non-degenerate Fermi surface enclosing a TRIM has a finite Chern number on each pocket and the TRIM becomes a Kramers Weyl point.
	
	For illustration, we apply strain on BiTeI to break all the mirror symmetries of the crystal. The compressive strain is achieved by reducing the lattice constant $\mathbf{a_1}$ of the crystal as shown in Fig.~\ref{fig:fig2}a. The evolution for the band structures along $\Gamma$--A under $1\%$, $3\%$ and $5\%$ strain strengths is summarized in Fig.~\ref{fig:fig4}c, \ref{fig:fig4}d and \ref{fig:fig4}e, respectively. (Note that the KNL $\Gamma$--A in the case without strain is shown in Fig.~\ref{fig:fig2}c.) Impressively, we found the KNL $\Gamma$--A in BiTeI  can be split sizably ($\sim$ order of tens of meV) by less than $3\%$ strain,  and the $\Gamma$ and A points become Kramers Weyl points with opposite chirality.  As A is the only Weyl point which is close to the Fermi energy while other Weyl points are at least 200 meV above, a single Weyl point near the Fermi energy is generated. Although there is only a single Weyl point near the Fermi energy,  the Nielsen-Ninomiya theorem is not violated because there are two Fermi pockets  carrying opposite chiral charges which enclose this Weyl point.    Therefore, straining achiral crystals provides a new way to create Kramers Weyl semimetals. In Fig.~\ref{fig:fig4}f, we demonstrate how the chiral charge $\mathcal{C}$  of this strain-induced Kramer Weyl point  can be measured by the circular photogalvanic effect \cite{Moore}. It is clear that when a Kramers Weyl point is created, the system shows the quantized circular photogalvanic effect. The details are given in the Supplementary Note 6.
	
	\section{Discussion}
	
	In this work, we point out that all non-centrosymmetric achiral crystals possess KNLs which connect TRIMs across the whole Brillouin zone. It is important to note that the KNLs are very different from nodal lines  generated by band inversions which can only be accessed in a very small range of energy window \cite{Burkov,Weng2,Fang3, Bian}. As illustrated in the band structure calculation of Fig.~\ref{fig:fig2}d and \ref{fig:fig2}j, KNLs appear in all the bands connecting some TRIMs. These KNLs create the  spindle torus type and the octdong type Fermi surface as long as the Fermi surfaces enclose TRIMs at an arbitrary Fermi energy. As listed in Table~\ref{table1}, a large number of existing materials are indeed KNLMs. Moreover, generic nodal lines formed by band inversion \cite{Fang_2016}  can be removed without breaking any symmetries. In sharp contrast, the KNLs are enforced and protected by a combination of the time-reversal symmetry and achiral crystal symmetries. The KNLs cannot be removed unless these symmetries are broken.
	
	Here, we briefly discuss some other possible physical consequences of KNLMs when the KNLs are gapped out. One way to gap out the KNLs is by shining a circularly polarized light on the material, which breaks time-reversal symmetry and in principle can lift the degeneracy of KNLs. This can result in sizable  Berry curvature around the KNLs and lead to  a light-induced  anomalous Hall effect as in the case of   graphene \cite{McIver}, where  anomalous Hall current arises due to the finite Berry curvature from the light-induced gapped Dirac cone. However, due to the large number of two-dimensional massless Dirac fermions in the material, we expect the effect is larger than that in graphene. Another  possibility  is to gap out the KNL through a Zeeman field, which can give rise to a  field-induced anomalous Hall effect.
	
	So far, we have only discussed KNLs in symmorphic crystals in detail. Indeed, KNLs also appear in all crystals that are non-centrosymmetric and nonsymmorphic. Particularly, there are always KNLs coming out of the $\Gamma$ points of nonsymmorphic crystals. Therefore, we conclude that all non-centrosymmetric achiral crystals possess KNLs, which is the central result of this work.  However, the situations in nonsymmorphic crystals are more complicated. For example, as discussed in Supplementary Note 7, screw symmetries can enforce nodal planes at Brillouin boundaries  which overwhelm the KNLs in these planes, while  glide mirror symmetries can enforce KNLs that are perpendicular to the glide mirror plane at Brillouin zone boundaries. Furthermore, bands at TRIM with higher-fold (such as four-fold and eight-fold) degeneracy are widely supported in nonsymmorphic crystals. For example,  the TRIM R in  nonsymmorphic SG No.~218 ($P\bar{4}3n$) and the TRIM H in  nonsymmorphic SG No.~220 ($I\bar{4}3d$) allows eight-dimensional corepresentations, which is consistent with the work of Wieder et al. \cite{Wieder} and Bradlyn et al. \cite{Bradlyn2}. As excepted, in these cases, the KNLs still emerge from these achiral TRIMs as shown in Supplementary Note 7. Specifically, the eight-fold degeneracies at the TRIM H in SG.~220 ($I\bar{4}3d$) split into four non-degenerate bands and two KNLs along H--P directions, or four KNLs along H--N and H--$\Gamma$ directions. However, a complete understanding of how the KNLs appear in nonsymmorphic  achiral crystals requires more study in the future.
	
	\section{Methods.}
	
	\begin{table*}
		\caption{The $\mathbf{k}\cdot \mathbf{p}$ Hamiltonians at TRIMs with noncentrosymmetric achiral little groups. The point group symmetry, the corresponding abstract group symbols together with time-reversal invariant irreducible corepresentations (IR coreps) are listed. The general form of the Hamiltonians and the direction of the KNLs are listed. In general, the KNLs lie along some high symmetry directions such as the $z$-direction. For points groups $C_{1v}$, $C_{3v}$ and $C_{3h}$, the KNLs lie within the mirror planes which is denoted as $\in m$.  Here $k_{\pm}=k_x\pm ik_y$, the Pauli matrices $\sigma_{x,y,z}$ operate on the spinor basis with $J_z=\pm 1/2$ or $J_z= \pm 3/2$, and $\hat{J}_i$ are the angular momentum operators with $J=3/2$.}\label{table2}
		\begin{tabular}{ccccc}
			\hline\hline
			Point group & IR coreps \cite{Bradley}  & $d$& $\mathbf{k}\cdot\mathbf{p}$ Hamiltonian& Directions of KNLs \\\hline
			$C_{1v}$& $G_4^1: R_2R_4$& 2&$\alpha_{13}k_z\sigma_x+\alpha_{23}k_z\sigma_y+(\alpha_{31}k_x+\alpha_{32}k_y)\sigma_z$&$\in m$\\
			$C_{2v}$& $G_8^5: R_5$&2&$\alpha_{12}k_y\sigma_x+\alpha_{21}k_x\sigma_y$&$\hat{z}$\\
			$S_4$&$G_8^1: R_2R_8, R_4R_6$&2& $(\alpha_{11}k_x+\alpha_{12}k_y)\sigma_x+(\alpha_{12}k_x-\alpha_{11}k_y)\sigma_y$&$\hat{z}$\\
			$C_{4v}$& $G_{16}^{14}: R_6, R_7$&2&$\alpha_{12}k_y\sigma_x-\alpha_{12}k_x\sigma_y$&$\hat{z}$\\
			$D_{2d}$&$G_{16}^{14}: R_6, R_7 $&2&$\alpha_{11}k_x\sigma_x-\alpha_{11}k_y\sigma_y$&$\hat{z}$\\
			$C_{3v}$&$G_{12}^4: R_3R_4$&2&$i\alpha_1 (k_{+}^3-k_{-}^3)\sigma_x+(\alpha_2k_z^3+\alpha_4(k_{+}^3+k_{-}^3))\sigma_y+i\alpha_5 (k_{+}^3-k_{-}^3)\sigma_z$&$ \in m$\\
			&$G_{12}^4: R_6$&	2&$\alpha_{12}k_y\sigma_x-\alpha_{12}k_x\sigma_y$& $\hat{z}$\\
			$C_{3h}$&$G_{12}^1: R_4R_{10}, R_6R_8$&2&$(\beta_1 k_{+}^2+\beta_1^*k_{-}^2)k_z\sigma_x+i(\beta_1k_{+}^2-\beta_{1}^*k_{-}^2)k_z\sigma_y+(\beta_2k_{+}^3+\beta_2^*k_{-}^3)\sigma_z$& $\hat{z},\in m$\\
			&$G_{12}^1: R_2R_{12}$&2&$(\alpha_1 k_{z}^3+\alpha_2k_{+}k_{-}k_z)\sigma_x+(\alpha_3k_z^3+\alpha_4k_{+}k_{-}k_z)\sigma_y+(\beta_1k_{+}^3+\beta_1^*k_{-}^3)\sigma_z$&$\in m$\\
			$C_{6v}$&$G_{24}^{11}: R_7, R_8$&2&$\alpha_{12}k_y\sigma_x-\alpha_{12}k_x\sigma_y$&$\hat{z}$\\
			&$G_{24}^{11}: R_9$&2&$i\alpha_1 (k_{+}^3-k_{-}^3)\sigma_x+\alpha_2(k_{+}^3+k_{-}^3)\sigma_y$&$\hat{z}$\\
			$D_{3h}$&$G_{24}^{11}: R_7, R_8$&2&$(\alpha_1 k_{z}^3+\alpha_2k_{+}k_{-}k_z)\sigma_y+i\alpha_3(k_+^3-k_-^3)\sigma_z$&$\hat{x}$, $C_3\hat{x}$,$C^2_3\hat{x}$, $\hat{z}$\\
			&$G_{24}^{11}: R_9$&2&$(\alpha_1 k_{z}^3+\alpha_2k_{+}k_{-}k_z)\sigma_y++i\alpha_3(k_+^3-k_-^3)\sigma_z$& $\hat{x}$, $C_3\hat{x}$,$C^2_3\hat{x}$\\
			$T_d$&$G_{48}^{10}: R_4, R_5$&2&$\alpha(k_x(k_y^2-k_z^2)\sigma_x+k_y(k_z^2-k_x^2)\sigma_y+k_z(k_x^2-k_y^2)\sigma_z)$& $\hat{x},\hat{y},\hat{z}, \pm\hat{x}\pm\hat{y}\pm\hat{z}  $ \\
			&$G_{48}^{10}: R_8$&4&$\beta \sum_{i} k_i^2\hat{J}_i^2+\gamma\sum_{i\neq j} k_ik_j\hat{J}_i\hat{J}_j+\delta\sum_ik_i(\hat{J}_{i+1}\hat{J}_i\hat{J}_{i+1}-\hat{J}_{i+2}\hat{J}_i\hat{J}_{i+2})$& $\hat{x},\hat{y},\hat{z}, \pm\hat{x}\pm\hat{y}\pm\hat{z}$\\\hline
		\end{tabular}
	\end{table*}

	{\bf $\mathbf{k}\cdot\mathbf{p}$ Hamiltonians  near TRIMs with achiral little group symmetry } In this section, we provide the general forms of the $\mathbf{k}\cdot\mathbf{p}$ Hamiltonians near the TRIM points of symmorphic crystals with achiral little group symmetry to help to understand how KNLs emerge from TRIMs, as listed in Table~\ref{table2}. It is important to note that these $\mathbf{k}\cdot\mathbf{p}$ Hamiltonians can also describe the $\Gamma$ point of nonsymmorphic crystals.
	
	In Table~\ref{table2}, we enumerate all allowed irreducible corepresentations of the ten noncentrosymmetric achiral point groups, the corresponding $\mathbf{k}\cdot\mathbf{p}$ Hamiltonians as well as directions of KNLs.  
	Here, we use the convention given in Ref. \cite{Bradley} where the irreducible representations of abstract groups (AGs) are introduced, to label the time-reversal invariant corepresentations.
	To summarize, we note that: (1) There are doubly degenerate KNLs emerging from all TRIM points with achiral little group symmetry. (2) KNLs lie along high symmetry directions in most point groups except certain irreducible corepresentations in $C_{1v}$, $C_{3v}$ and $C_{3h}$, in which  cases the KNLs can be pinned along some generic directions within mirror-invariant planes as denoted by the symbol $\in m$. (3) All the irreducible corepresentations are two-dimensional except for the $T_d$ point group which allows a four-dimensional corepresentation. The general form of this four-dimensional Hamiltonian is expressed with $J_i$ which is the angular momentum operators of $J=3/2$ states, with $i =x,y,z$. It is important to note that there are doubly degenerate KNLs emerging from TRIMs with four-dimensional corepresentations.
	


	Next, we apply Table~\ref{table2} to understand the KNLs in the band structure of some realistic materials.
	In  BiTeI, the TRIMs $\Gamma$ and A respect $C_{3v}$ symmetry, which allows  time-reversal invariant corepresentations $G^{4}_{12}: R_3R_4$ and $G_{12}^4: R_6$. For energy bands at TRIMs described by  corepresentations $R_6$, the $\mathbf{k}\cdot \mathbf{p}$ Hamiltonian of the spin-orbit coupling term is
	\begin{equation}
	H_{so}(\mathbf{k})=	\alpha_{12}(k_y\sigma_x-k_x\sigma_y),  
	\end{equation}
	which allows a degenerate line along $\hat{z}$ direction as listed in Table~\ref{table2} and explains the KNL $\Gamma$--A in Fig.~\ref{fig:fig2}c. Similarly, in Cr$_2$AgBiO$_8$,  the TRIMs $\Gamma$ and $Z$ respect $S_4$ symmetry, and the corresponding time-reversal invariant  irreducible corepresentations are $G_8^1$: $R_2R_8$ and $R_4R_6$. For energy bands at TRIMs described by these corepresentations, the $\mathbf{k}\cdot \mathbf{p}$ Hamiltonian of spin-orbit coupling term is
	\begin{equation}
	H_{so}(\mathbf{k})=	(\alpha_{11}k_x+\alpha_{12}k_y)\sigma_x+(\alpha_{12}k_x-\alpha_{11}k_y)\sigma_y,  
	\end{equation}
	which vanishes along $\hat{z}$ direction and is consistent with the support of KNL $\Gamma$--Z shown in Fig.~\ref{fig:fig2}g. 
	
	As shown in Fig.~\ref{fig:fig2}d and Fig.~\ref{fig:fig2}j, there are KNLs lying within the mirror plane when TRIMs respect $C_{1v}$ symmetry. This property is also manifested by the $\mathbf{k}\cdot \mathbf{p}$ Hamiltonian.  The Hamiltonians near such TRIMs have the form 
	\begin{equation}
	H_{so}(\mathbf{k})=\alpha_{13}k_z\sigma_x+\alpha_{23}k_z\sigma_y+(\alpha_{31}k_x+\alpha_{32}k_y)\sigma_z,\label{kpc1v}
	\end{equation}  
	where the mirror operation is  $m_z: z\mapsto -z$. Evidently, the $H_{so}(\mathbf{k})$ vanishes  along $(-\alpha_{32}, \alpha_{31},0)$, which is a direction within the mirror plane. 
	
	In the previous sections, we focused on the KNLs which emerge from two-fold degenerate points at TRIMs. However, we note that the $T_d$ point group allows a four-dimensional irreducible corepresentations $G_{48}^{10}: R_8$. The  $\mathbf{k}\cdot \mathbf{p}$ Hamiltonian in basis spanned by states with total angular momentum $J=3/2$ and azimuthal quantum number $J_z$ (i.e. $\ket{3/2,J_z}$  with $J_z=\pm 3/2,\pm1/2$) can be written as \cite{Agterberg}
	\begin{align}
	H_{so}(\mathbf{k})&=\beta \sum_{i} k_i^2\hat{J}_i^2+\gamma\sum_{i\neq j} k_ik_j\hat{J}_i\hat{J}_j\nonumber\\
	&+\delta\sum_ik_i(\hat{J}_{i+1}\hat{J}_i\hat{J}_{i+1}-\hat{J}_{i+2}\hat{J}_i\hat{J}_{i+2}),
	\label{Eq:half_Hu}
	\end{align}
	where $i=x,y,z$ and $i+1=y$ if $i=x$, etc.  $\hat{J}_i$ are the $4\times 4$ matrices of the $J=3/2$ angular momentum operators.
	This Hamiltonian results in KNLs along $\hat{x},\hat{y},\hat{z}$ and $\pm\hat{x}\pm\hat{y}\pm\hat{z}$. This is consistent with the KNLs found in HgSe (SG No. 216, $F\bar{4}3m$), YPtBi (SG No. 216, $F\bar{4}3m$) as shown in the Supplementary Note 7. It can be seen from the band structure calculations that the four-dimensional corepresentations are decomposed into two two-dimensional irreducible representations along $\Gamma$--X and one two-dimensional irreducible representation plus two one-dimensional representations along $\Gamma$--L.

	\section*{Acknowledgments}	
	The authors thank the discussions with Zhijun Wang, Quansheng Wu, Andrei Bernevig, Xi Dai, Joel Moore, Titus Neupert, Adrian Po and Binghai Yan. KTL acknowledges the support of the Croucher Foundation, the Dr. Tai-chin Lo Foundation and the HKRGC through grants C6025-19G, RFS2021-6S03, 16310219, 16309718 and 16310520.
	
	\section*{Author Contributions}
	K.T.L. conceived the idea of Kramers nodal lines and supervised the project. Y.-M.X., X.-J.G.  performed the major part of the calculations and material analysis. Y.-M.X., X.-J.G. and K.T.L. wrote the manuscript with contributions from all authors. X.Y.X. performed the DFT calculations for BiTeI. C.-P.Z., J.-X.H. and J.Z.G. contributed to part of the calculations and are involved in discussions.
	
	\section*{Competing Interests}
	The authors declare no competing interests.
	
	\section*{Data availability}
	The data that support the findings of this study are available from the corresponding author upon reasonable request.
	
	\section*{Code availability}
	The computer codes that support the findings of this study are available from the corresponding author upon reasonable request.

\clearpage

\onecolumngrid
\begin{center}
	\textbf{\large Supplementary Material for  \lq\lq{}Kramers Nodal line Metals\rq\rq{}}\\[.2cm]
	Ying-Ming Xie,$^{1}$ Xue-Jian Gao,$^{1}$, Xiao Yan Xu,$^2$, Cheng-Ping Zhang$^1$, Jin-Xin Hu$^1$, Jason Z. Gao$^1$ and K. T. Law$^{1,*}$\\[.1cm]
	{\itshape ${}^1$Department of Physics, Hong Kong University of Science and Technology, Clear Water Bay, Hong Kong, China}\\
	{\itshape ${}^2$Department of Physics, University of California at San Diego, La Jolla, California 92093, USA}\\[1cm]
\end{center}
\setcounter{equation}{0}
\setcounter{section}{0}
\setcounter{figure}{0}
\setcounter{table}{0}
\setcounter{page}{1}
\renewcommand{\theequation}{\arabic{equation}}
\renewcommand{\thesection}{Supplementary Note \arabic{section}}
\renewcommand{\figurename}{Supplementary Figure}
\renewcommand{\thetable}{\arabic{table}}
\renewcommand{\tablename}{Supplementary Table}
\makeatletter

\twocolumngrid

	\maketitle
\section{DFT calculations} Throughout this work, the Vienna Ab initio Simulation Package (VASP) \cite{Kresse} with the projector-augmented wave method \cite{Blochl} and the Perdew-Berke-Ernzerhof’s (PBE) exchange-correlation functional in the generalized-gradient approximation \cite{Perdew,Langreth} was used to perform the first-principles density functional theory (DFT) calculations \cite{Hohenberg}. Information about calculated materials such as the lattice structures was obtained mainly from several material databases, \textit{e.g.}  the Materials Project \cite{Jain}, the Topological Materials Database \cite{Hinuma}, the Inorganic Crystal Structure Database (ICSD) \cite{ISCD} and the  TopoMat Database \cite{Topomat}.

To further look into the topological properties as well as to plot the special spindle torus and octdong Fermi surfaces of KNLMs, maximally localized generalized Wannier bands of some KNLMs (such as BiTeI, HgSe, Bi$_2$Te$_2$Br$_2$O$_9$ and Cr$_2$AgBiO$_8$) were projected from the first-principles results through the Wannier90 package \cite{Marzari,Mostofi} linked to VASP.

The open-source package WannierTools was used for post-processing of the Wannier tight-binding Hamiltonian \cite{WuQ}. These processes include the Fermi surface plotting, the calculation of Fermi arcs, surface states and chiral charges of Kramers Weyl points.

\section{A general proof of the existence of KNLs in achiral crystals}

\subsection{Notations}

The space group $G$ consists of all operations $\{R_{\alpha}|\mathbf{t}\}$ which leave a given lattice invariant, where the space group operator $\{R_{\alpha}|\mathbf{t}\}\mathbf{r}=\mathbf{r'}=R_{\alpha}\mathbf{r}+\mathbf{t}$.  A space group is called symmorphic  if there is a point such that all symmetries are the product of a symmetry fixing this point and a translation. For symmorphic space groups, the point group $\mathcal{G}$ is isomorphic to the factor group $G/T$ with $T$ as the translational group forming by translational operations $\{E|\mathbf{t}\}$ that leave the lattice to be invariant, $E$ is identity operation. On the contrary, nonsymmorphic space groups cannot be represented as semi-direct product groups of a discrete translation group $T$ and a corresponding point group $\mathcal{G}$.

The little group $G_{\mathbf{k}}$ of a wave vector  $\mathbf{k}$ is formed by the set of space group operations $\{R_{\alpha}|\mathbf{t}\}$ such that $R_{\alpha}\mathbf{k}=\mathbf{k}+\mathbf{G}_i$, $\mathbf{G}_i$ is the reciprocal lattice vector.  For our purpose, it is sufficient to determine the reps of Herring's little group $^HG^{\mathbf{k}}=G^{\mathbf{k}}/T^{\mathbf{k}}$, where $T^{\mathbf{k}}$ is the group of the translational symmetry operations $\{E|\mathbf{t}\}$ with $\exp(-i\mathbf{k}\cdot \mathbf{t})=1$.  The $^HG^{\mathbf{k}}$ in general can be identified with one of the abstract groups (AGs) given in \cite{Bradley_book}. For a symmorphic space group, the Herring's little group $^HG^{\mathbf{k}}$ is always isomorphic to a point group $\mathcal{G}_{\mathbf{k}}$.  Throughout this work, the little group refers to the Herring's little group, where the integer translations have been factored away. 


When there exhibits an additional anti-unitary symmetry $\mathcal{T}$ with $\mathcal{T}^2=-1$, such as time-reversal symmetry, that leaves $\mathbf{k}$ to be invariant, the symmetry group becomes $G_{\mathbf{k}}+\mathcal{T}G_{\mathbf{k}}$. In this case, the states at TRIMs are described by the corepresentations of the symmetry group $G_{\mathbf{k}}+\mathcal{T}G_{\mathbf{k}}$. And since the translational operations  is not essential here, we can directly use the corepresentations of the Herring's little group $^HG^{\mathbf{k}}$ to label the states at TRIMs. Furthermore, according to the theory of corepresentations \cite{Bradley_book}, if a corepresentation $D^{\Gamma_i}$ is real or complex, the corep is irreducible and the degeneracies at TRIMs will be doubled due to the anti-unitary symmetry, while if $D^{\Gamma_i}$ is pseudo-real, the corepresentation becomes reducible and there is no extra degeneracy from this anti-unitary symmetry. A more systematic introduction to corepresentations can be found in Ref. \cite{Bradley_book}.

\subsection{Symmetry properties of the SOC term}

The bands near a TRIM $\mathbf{k}_0$ with a two-fold degeneracy can be described by a two-band  Hamiltonian 
\begin{equation}
H(\mathbf{k})=f_0(\mathbf{k})\sigma_0+ \bm{f}(\mathbf{k})\cdot\mathbf{\sigma},
\label{eq:1}
\end{equation}
where $\bm{f}$($\mathbf{k}$)$\cdot\mathbf{\sigma}$ denotes the spin-orbit coupling term (SOC), and $\mathbf{\sigma}$ are Pauli  matrices operating on spin space $\ket{\pm \frac{1}{2}}$.  This Hamiltonian needs to respect the symmetry $\mathcal{T}\times \mathcal{G}_{\mathbf{k}_0}$, where $\mathcal{T}=i\sigma_y K$ with $K$ as complex conjugate is time-reversal symmetry, and $\mathcal{G}_{\mathbf{k}_0}$ is the point group symmetry that the Herring's little group $^HG^{\mathbf{k}}$ is isomorphic to. The time-reversal symmetry requires $\bm{f}(\mathbf{k})=-\bm{f}(-\mathbf{k})$, while the constraint imposed by a symmetry operation $R$ in $\mathcal{G}_{\mathbf{k}_0}$ is $H(\mathbf{k})=U^{-1}_{1/2}(R)H(R\mathbf{k})U_{1/2}(R)$, \textit{i.e.}, 
\begin{align}
\bm{f}(\mathbf{k})\cdot\mathbf{\sigma}&=U^{-1}_{1/2}(R)\bm{f}(R\mathbf{k})\cdot\mathbf{\sigma}U_{1/2}(R)\nonumber\\
&=\text{Det}(R)\bm{f}(R\mathbf{k})\cdot (R\mathbf{\sigma})\nonumber\\
&=\text{Det}(R) R^{-1}\bm{f}(R\mathbf{k})\cdot \mathbf{\sigma}.
\end{align}
\begin{equation}
\bm{f}(\mathbf{k})=\text{Det}(R)R^{-1}\bm{f}(R\mathbf{k}), \label{transf}
\end{equation} 
where $R\in O(3)$, and $U_{1/2}(R)$ is the $SU(2)$ representation of $R$.  

\subsection{Symmetry transformation properties of the linear term}
When $\bm{f}(\mathbf{k})$ is dominant by linear terms, $\bm{f}(\mathbf{k})$ can be written as
\begin{equation}
\bm{f}(\mathbf{k})=\hat{M}\mathbf{k},
\end{equation}
where $\hat{M}$ is a 3-by-3 matrix. According to Supplementary Eq.~(\ref{transf}),
\begin{equation}
\hat{M}\mathbf{k}=\text{Det}(R)R^{-1}\hat{M}R\mathbf{k}. 
\end{equation}
Hence
\begin{equation}
\hat{M}=\text{Det}(R)R^{-1}\hat{M}R,\label{matm}
\end{equation}
and
\begin{equation}
\text{Det}(\hat{M})=\text{Det}(R)\text{Det}(\hat{M}).
\end{equation}
For achiral point groups $\mathcal{G}_{\mathbf{k}_0}$, there exists a roto-inversion operation $\tilde{R}$ with $\text{Det}(\tilde{R})=-1$, which further requires $\text{Det}(\hat{M})=0$. Therefore, in an achiral point group, the determinant of $\hat{M}$ is always zero.

In the main text, we have assumed that the matrix $\hat{M}$ of an achiral point group is always diagonalizable, which can be verified by enumerating all possible forms of $\hat{M}$ for different achiral point groups (Supplementary Table~\ref{Table_S1}). However, in some cases ($C_{3v}$, $C_{4v}$ and $C_{6v}$), not all the eigen-values $\epsilon_j$ or eigen-vectors $\mathbf{n_j}$ of $\hat{M}$ are real. In spite of this fact, our argument in the main text still holds as the null eigen-vector $\mathbf{n_3}$ with eigen-value zero is always a real vector (multiplied by an overall trivial phase). This can be easily proved considering $\hat{M}$ is a real matrix.

\subsection{KNLs enforced by roto-inversion symmetries: mirror, $S_3$ and $S_4$ symmetry}

Let us further study the constraint of roto-inversion ($\text{Det}(\tilde{R})=-1$) on the specific form of $\bm{f}(\mathbf{k})$. For convenience, we use $k_{1,2}$ and $k_{3}$ to denote the coordinates perpendicular and parallel to the roto-inversion axis respectively. In general, a roto-inversion operation can be decomposed into a combination of an inversion $I$ and a rotation $C_n$, \textit{i.e.}
\begin{equation}\label{Eq:S8}
\tilde{R} = I \cdot C_n
\end{equation}
Following Supplementary Eq.~(\ref{transf}), the constraints of time-reversal symmetry and this roto-inversion symmetry impose
\begin{widetext}
	\begin{eqnarray}
	f_{\pm}(k_{+},k_{-},k_3)=-f_{\pm}(-k_{+},-k_{-},-k_3)=e^{\mp i\varphi}f_{\pm}(-e^{+i\varphi} k_{+}, -e^{-i\varphi}k_{-},-k_{3}),\label{Eq:f_pm}\\
	f_{3}(k_{+},k_{-},k_3)=-f_{3}(-k_{+},-k_{-},-k_3)=f_{3}(-e^{+i\varphi} k_{+}, -e^{-i\varphi}k_{-},-k_{3}).\label{Eq:f3}
	\end{eqnarray}
\end{widetext}
Here, $f_{\pm}=f_1\pm if_2$, $k_{\pm}=k_1\pm ik_2$, and $\varphi=2\pi/n$. With $f_{\pm}(\mathbf{k})$,  the eigen-energies of $H(\mathbf{k})$ can be written as $E_\pm(\mathbf{k})=f_0(\mathbf{k})\pm\sqrt{f_{+}(\mathbf{k})f_{-}(\mathbf{k})+f_3(\mathbf{k})^2}$. We should note here that the origin point of $\mathbf{k}$ vector in Supplementary Eq.~(\ref{Eq:f_pm}) and Supplementary Eq.~(\ref{Eq:f3}) is not necessary to be $\Gamma$ but any TRIM with an achiral little group.
In the following, we  show that all roto-inversion symmetries mirror, $S_3$, $S_4$ enforce KNLs. (Note there does not contain $S_6$ in non-centrosymmetric achiral point group), where the $S_n$ symmetry is defined as
\begin{equation}
S_n=m\cdot C_n,
\end{equation} 	
representing the combination of a mirror and  a $n$-fold  rotation perpendicular to the mirror plane according to the Schoenflies notation. As a result, in Supplementary Eq.~(\ref{Eq:S8}), $n=6$ for $S_3$ and $n=4$ for $S_4$.

(I) \textit{For an achiral crystal with  mirror symmetry,  there always exist KNLs within the mirror plane}.

\begin{figure}
	\centering
	\includegraphics[width=1\linewidth]{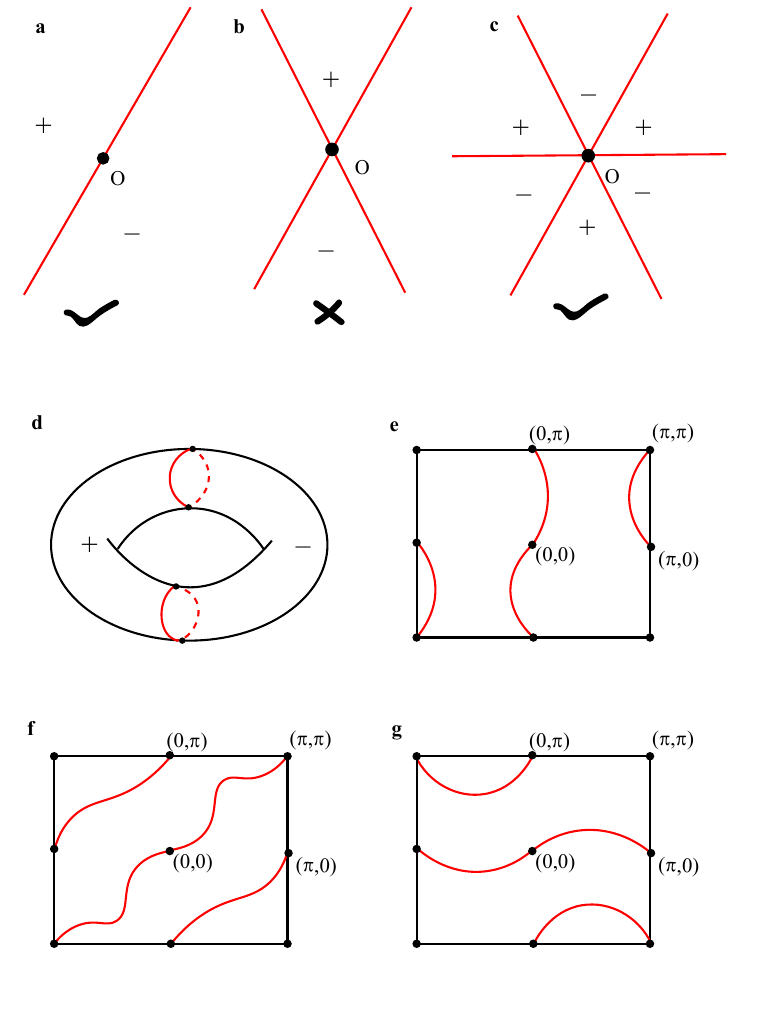}
	\caption{KNLs within the mirror plane. {\bf a}, {\bf b}, {\bf c} schematic plots of degenerate lines (red lines) coming from a TRIM  within the mirror plane. $\pm$ here labels the sign of the scalar function $f_3(k_{\parallel})$. 
		{\bf d} KNLs must completely cut the mirror-invariant Brillouin plane, which is topologically equivalent to a torus surface, into two separate parts. This requirement implies that there should be at least two KNLs within one mirror-invariant $k$ plane. {\bf e}, {\bf f} and {\bf g}  Schematics for the three skeleton cases for KNLs within the mirror-invariant Brillouin plane, which are guaranteed by the time-reversal and a mirror symmetry.}
	\label{fig:figs1}
\end{figure}

Before starting proceeding this part, we need to state two facts that if a crystal respects mirror symmetry $m$: (i) the set of primitive reciprocal lattice vectors can always be chosen in such a way that exactly two of them lies within any pre-chosen $m$-invariant $k$-plane. (ii) The $m$-invariant $k$-plane contains exactly four non-equivalent TRIMs, though some of them need not lie within the 1st Brillouin zone.

For a mirror symmetry ($n=2$, $\varphi=\pi$, Supplementary Eq.~(\ref{Eq:f_pm}) and Supplementary Eq.~(\ref{Eq:f3}) yield
\begin{eqnarray}
f_{\pm}(k_{\parallel},k_3)=-f_{\pm}(-k_{\parallel},-k_3)=-f_{\pm}(k_{\parallel},-k_3),\\
f_{3}(k_{\parallel},k_3)=-f_{3}(-k_{\parallel},-k_3)=f_{3}(k_{\parallel},-k_3).
\end{eqnarray}
where $k_{\parallel}=(k_1,k_2)$. Thus on the mirror-invariant $k$-planes where $k_3=0$ or $\pi$,  $f_\pm$ terms vanish and the only finite $f_3$ term is odd in $k_\parallel$, \textit{i.e.} $f_{3}(k_{\parallel})=-f_{3}(-k_{\parallel})$. Note any TRIM lying on the plane can be chosen as the origin point of $k_\parallel$.

Now the degenerate lines upon this plane are given by the equation
\begin{equation}
f_{3}(k_{1},k_{2})=0.
\end{equation}
And importantly, $f_{3}(k_1, k_2)$ is an odd (relative to any TRIM) scalar function defined on a 2D torus $k$-surface. Globally speaking, due to the odd function behavior of $f_3$, there exists at least one positive-valued area and correspondingly one negative-valued area, the boundaries of which give the KNLs and must pass through the TRIMs.   KNLs emerging from TRIMs   are thus actually  protected by the odd property of $f_3$ as well as the topology of the $k$-surface. As the boundaries splitting the positive and negative areas, these KNLs have the following properties: (i) they have no end points; (ii) for a given TRIM, only an odd number of KNLs can come out (Supplementary Fig.~\ref{fig:figs1}{\bf a}$\sim${\bf c}); 
(iii) they must cut the torus $k$-surface into at least two separate parts, which implies that there are at least two KNLs for a mirror-invariant $k$-plane ((Supplementary Fig.~\ref{fig:figs1}{\bf d}); (iv) they must connect two TRIMs. A time-reversal symmetric closed loop on the torus has to pass through  two TRIMs (one is $k_{\parallel}=0$, and the other is $k_{\parallel}\equiv -k_{\parallel}$ at the Brillouin zone boundary). Therefore, in the mirror plane, each degenerate line  coming out from one TRIM  has to connect with another TRIM, which forms the KNLs. Considering all these properties of the KNLs, the three simplest cases of how they should exist on the mirror-invariant $k$-plane is given in Supplementary Fig.~\ref{fig:figs1}{\bf e}$\sim${\bf g}, and all other more complicated cases are generated by adding more nodal lines to these three skeleton cases.

As shown above, the time-reversal and a mirror symmetry pin the KNLs upon the mirror-invariant $k$-plane, while allowing their tracks to go along quite arbitrary curves on the plane. However, additional crystal symmetries like $C_2$ and $C_3$ rotations can further constrain KNLs along some high-symmetry paths. We take 
a simple case as an illustration where the additional symmetry is a $C_2$ rotation whose rotation axis is within the mirror plane. Without loss of generality, the axis of $C_2$ can be set along the $k_x$-direction, \textit{i.e.} $C_{2x}$. This new $C_{2x}$ rotation requires the form of odd function $f_3(k_x, k_y)$ further satisfying $f_3(k_x, -k_y)=-f_3(k_x,k_y)$. Along the $C_{2x}$ axis, $f_3(k_x,k_y=0)$ vanishes and give rise to a straight KNL joining two TRIMs. Other additional symmetry cases can also be analyzed in this way, but a more systematic method to find all high-symmetry KNLs is to utilize the compatibility relation as we have presented in the main text.


(II) \textit{ For an achiral crystal with  roto-inversion symmetry $S_3(n=6)$ or $S_4(n=4)$, there always exists a KNL along the $\langle0,0,k_3\rangle$-direction which is perpendicular to the roto-inversion plane.}

Along the $\langle0,0,k_3\rangle$-direction,  Supplementary Eq.~(\ref{Eq:f_pm}) and Supplementary Eq.~(\ref{Eq:f3}) are simplified as
\begin{eqnarray}
&f_{\pm}(k_{3})=-f_{\pm}(-k_{3})=e^{\mp i\varphi}f_{\pm}(-k_{3}),\label{eq:on_axis} \\
&f_{3}(k_{3})=-f_{3}(-k_{3})=f_{3}(-k_{3})
\end{eqnarray}
Hence, $f_{\pm}(\mathbf{k}), f_{\pm}(\mathbf{k})$ and $f_{3}(k_{3})$  must vanish along the $\langle0,0,k_3\rangle$-direction when $(1+e^{\mp i\varphi})\neq0$, which is the case for $S_3$ symmetry ($\varphi=\pi/3$) and $S_4$ symmetry ($\varphi=\pi/2$). In contrast for mirror symmetries with $\varphi=\pi$, we have $(1+e^{\mp i\varphi})=0$, which allows finite $f_\pm$ that results in a finite splitting along the $\langle0,0,k_3\rangle$-direction.

\section{ KNLs from achiral little groups based on k$\cdot$ p analysis}

\begin{table*}
	\caption{ Kramers nodal lines (KNLs) from  TRIMs with achiral little groups based on {\bf k} $\cdot$ {\bf p} Hamiltonian analysis. Here the point group denotes the one that $^HG^{\mathbf{k}}$ is isomorphic to,  $k_{\pm}=k_x\pm ik_y$, and the Pauli matrices $\sigma$ in the Table operate on the corresponding basis.  }
	\begin{ruledtabular}
		\begin{tabular}{cccccccc}\label{Table_S1}
			Point group& P-axes&Coreps& Basis& {\bf k} $\cdot$ {\bf p} Hamiltonian& Matrix $\hat{M}$& KNL& Touching\\\hline
			&&&&&&&\\
			\multirow{3}{*}{$C_{1v}$}& \multirow{3}{*}{$\hat{z}$ ($m\perp \hat{z}$)}& \multirow{3}{*}{$G_4^1: R_2R_4$}& \multirow{3}{*}{$\ket{1/2,\pm 1/2}$}&$\alpha_{13}k_z\sigma_x+\alpha_{23}k_z\sigma_y+$ &\multirow{3}{*}{$\begin{pmatrix}
				0&0&\alpha_{13}\\
				0&0&\alpha_{23}\\
				\alpha_{31}&\alpha_{32}&0
				\end{pmatrix}$}&$\in m$&\multirow{3}{*}{linear}\\
			&&&&$(\alpha_{31}k_x+\alpha_{32}k_y)\sigma_z$&&($-\alpha_{32}\hat{x}+\alpha_{31}\hat{y}$)&\\
			&&&&&&&\\
			&&&&&&&\\
			\multirow{3}{*}{$C_{2v}$}&\multirow{3}{*}{$\hat{z}$}&\multirow{3}{*}{$G_8^5: R_5$}&\multirow{3}{*}{$\ket{1/2,\pm 1/2}$}&&\multirow{3}{*}{$\begin{pmatrix}
				0&\alpha_{12}&0\\
				\alpha_{21}&0&0\\
				0&0&0
				\end{pmatrix}$}&\multirow{3}{*}{$\hat{z}$}&\multirow{3}{*}{linear}\\
			&&&&$\alpha_{12}k_y\sigma_x+\alpha_{21}k_x\sigma_y$&&&\\
			&&&&&&&\\
			&&&&&&&\\
			\multirow{3}{*}{$S_{4}$}&\multirow{3}{*}{$\hat{z}$}&&&&\multirow{3}{*}{$\begin{pmatrix}
				\alpha_{11}&\alpha_{12}&0\\
				\alpha_{12}&-\alpha_{11}&0\\
				0&0&0
				\end{pmatrix}$}&\multirow{3}{*}{$\hat{z}$}&\multirow{3}{*}{linear}\\
			&&$G_8^1: R_2R_8$&$\ket{1/2,\pm 1/2}$&$(\alpha_{11}k_x+\alpha_{12}k_y)\sigma_x+$&&&\\
			&&$G_8^1: R_4R_6$&$\ket{3/2,\pm 1/2}$&$(\alpha_{12}k_x-\alpha_{11}k_y)\sigma_y$&&&\\
			&&&&&&&\\
			\multirow{3}{*}{$C_{4v}$}&\multirow{3}{*}{$\hat{z}$}&&&&\multirow{3}{*}{$\begin{pmatrix}
				0&	\alpha_{12}&0\\
				-\alpha_{12}&0&0\\
				0&0&0
				\end{pmatrix}$}&\multirow{3}{*}{$\hat{z}$}&\multirow{3}{*}{linear}\\
			&&$G_{16}^{14}: R_7$;&$\ket{1/2,\pm 1/2}$;&$\alpha_{12}k_y\sigma_x-\alpha_{12}k_x\sigma_y$&&&\\
			&&$G_{16}^{14}: R_6$&$\ket{3/2,\pm 1/2}$&&&&\\
			&&&&&&&\\
			\multirow{3}{*}{$D_{2d}$}&\multirow{3}{*}{$\hat{z}$}&&&&\multirow{3}{*}{$\begin{pmatrix}
				\alpha_{11}&0&0\\
				0&-\alpha_{11}&0\\
				0&0&0
				\end{pmatrix}$}&\multirow{3}{*}{$\hat{z}$}&\multirow{3}{*}{linear}\\
			&&$G_{16}^{14}: R_7$;&$\ket{1/2,\pm 1/2}$;&$\alpha_{11}k_x\sigma_x-\alpha_{11}k_y\sigma_y$&&&\\
			&&$G_{16}^{14}: R_6$&$\ket{3/2,\pm 1/2}$&&&&\\
			&&&&&&&\\
			\multirow{3}{*}{$C_{3v}$}&\multirow{3}{*}{$\hat{z}$}&\multirow{3}{*}{$G_{12}^4: R_6$}&\multirow{3}{*}{$\ket{1/2,\pm 1/2}$}&&\multirow{3}{*}{$\begin{pmatrix}
				0&\alpha_{12}&0\\
				-\alpha_{12}&0&0\\
				0&0&0
				\end{pmatrix}$}&\multirow{3}{*}{$\hat{z}$}&\multirow{3}{*}{linear}\\
			&&&&$\alpha_{12}k_y\sigma_x-\alpha_{12}k_x\sigma_y$&&&\\
			&&&&&&&\\
			&&&&&&&\\
			\multirow{3}{*}{$C_{6v}$}&\multirow{3}{*}{$\hat{z}$}&\multirow{3}{*}{$G_{24}^{11}: R_7, R_8$}&\multirow{3}{*}{$\ket{1/2,\pm 1/2}$}&&\multirow{3}{*}{$\begin{pmatrix}
				0&\alpha_{12}&0\\
				-\alpha_{12}&0&0\\
				0&0&0
				\end{pmatrix}$}&\multirow{3}{*}{$\hat{z}$}&\multirow{3}{*}{linear}\\
			&&&&$\alpha_{12}k_y\sigma_x-\alpha_{12}k_x\sigma_y$&&&\\
			&&&&&&&\\
			&&&&&&&\\\hline
			
			\multirow{3}{*}{$T_d$}&\multirow{3}{*}{ $\hat{x},\hat{y},\hat{z}$}&&&&\multirow{3}{*}{---}&&\multirow{3}{*}{linear}\\
			&&$G_{48}^{10}: R_4$;&$\ket{1/2,\pm 1/2}$;&$\alpha(k_x(k_y^2-k_z^2)\sigma_x+k_y(k_z^2-k_x^2)$&&$\hat{x}$, $\hat{y}$, $\hat{z}$&\\
			&&$G_{48}^{10}: R_5$&$\ket{3/2,\pm 1/2}$&$\sigma_y+k_z(k_x^2-k_y^2)\sigma_z)$&&$\pm\hat{x}\pm\hat{y}\pm\hat{z}$&\\
			&&&&&&&\\
			\multirow{3}{*}{$C_{3v}$}&\multirow{3}{*}{$\hat{z}$}&&\multirow{3}{*}{$\ket{3/2,\pm 3/2}$}&$i\alpha_1 (k_{+}^3-k_{-}^3)\sigma_x+(\alpha_2k_z^3+$&\multirow{3}{*}{---}&\multirow{3}{*}{$\in m$}&\multirow{3}{*}{linear}\\
			&&$G_{12}^4: R_3R_4$&&$\alpha_3k_+k_-k_z) +\alpha_4(k_{+}^3+k_{-}^3))\sigma_y$&&&\\
			&&&&$+i\alpha_5 (k_{+}^3-k_{-}^3)\sigma_z$&&&\\
			\multirow{3}{*}{$C_{3h}$}&\multirow{3}{*}{$\hat{z}$}&&&&\multirow{3}{*}{---}&\multirow{3}{*}{$\hat{z}$\& $\in m$}&\multirow{3}{*}{quadratic}\\
			&&$G_{12}^1: R_4R_{10}$;&$\ket{1/2,\pm 1/2}$;&$(\beta_1 k_{+}^2+\beta_1^*k_{-}^2)k_z\sigma_x+i(\beta_1k_{+}^2$&&\\
			&&$G_{12}^1: R_6R_{8}$&$\ket{3/2,\pm 1/2}$&$-\beta_{1}^*k_{-}^2)k_z\sigma_y+(\beta_2k_{+}^3+\beta_2^*k_{-}^3)\sigma_z$&&&\\
			&&\multirow{3}{*}{$G_{12}^1: R_2R_{12}$}&&&\multirow{3}{*}{---}&\multirow{3}{*}{ $\in m$}&\multirow{3}{*}{linear}\\
			&&&$\ket{3/2,\pm 3/2}$&$(\alpha_1 k_{z}^3+\alpha_2k_{+}k_{-}k_z)\sigma_x+(\alpha_3k_z^3$&&\\
			&&&&$+\alpha_4k_{+}k_{-}k_z)\sigma_y+(\beta_1k_{+}^3+\beta_1^*k_{-}^3)\sigma_z$&&&\\
			\multirow{3}{*}{$D_{3h}$}&\multirow{3}{*}{$\hat{z}$}&&&&\multirow{3}{*}{---}&\multirow{3}{*}{$\hat{x}$, $C_3\hat{x}$,$C^2_3\hat{x}$, $\hat{z}$}&\multirow{3}{*}{quadratic}\\
			&&$G_{24}^{11}: R_7$;&$\ket{1/2,\pm 1/2}$;&$i\alpha_1(k_+^2-k_{-}^2)k_z\sigma_x-\alpha_1(k_{+}^2$&&\\
			&&$G_{24}^{11}: R_8$&$\ket{3/2,\pm 1/2}$&$+k_{-}^2)k_z\sigma_y+i\alpha_2(k_{+}^3-k_{-}^3)\sigma_z$&&&\\
			&&\multirow{3}{*}{$G_{24}^{11}: R_9$}&&&\multirow{3}{*}{---}&\multirow{3}{*}{$\hat{x}$, $C_3\hat{x}$,$C^2_3\hat{x}$}&\multirow{3}{*}{linear}\\
			&&&$\ket{3/2,\pm 3/2}$&$(\alpha_1 k_{z}^3+\alpha_2k_{+}k_{-}k_z)\sigma_y+$&&\\
			&&&&$+i\alpha_3(k_+^3-k_-^3)\sigma_z$&&&\\
			&&&&&&&\\
			\multirow{3}{*}{$C_{6v}$}&\multirow{3}{*}{$\hat{z}$}&&\multirow{3}{*}{$\ket{3/2,\pm 3/2}$}&$i\alpha_1 (k_{+}^3-k_{-}^3)\sigma_x+$&\multirow{3}{*}{---}&\multirow{3}{*}{$\hat{z}$}&\multirow{3}{*}{cubic}\\
			&&$G_{24}^{11}: R_9$&&$+\alpha_2(k_{+}^3+k_{-}^3)\sigma_y$&&&\\
			&&&&&&&\\
		\end{tabular}
	\end{ruledtabular}
\end{table*}

To show the directions of KNLs coming from one TRIM explicitly, we derived the $\mathbf{k}\cdot\mathbf{p}$ Hamiltonians given by two dimensional double-valued irreducible representations (IRRs) for all non-centrosymmetric achiral little groups. In the Method Section of the main text, we have summarized these $\mathbf{k}\cdot \mathbf{p}$ Hamiltonians in Main text Table 2. Here, we summary Supplementary Table~\ref{Table_S1} to present more details  including the principle axes, the specific bases. As shown in Supplementary Table~\ref{Table_S1}, we further identified the touching types of KNLs given by each  $\mathbf{k}\cdot \mathbf{p}$ Hamiltonian.

Generally, there are always KNLs emerging from TRIMs  with achiral little groups. Notably, the features of KNLs of $J_z=\pm1/2$ and $J_z=\pm3/2$ fermions are different, where $J_z$ is the $z$ component of the total angular momentum.  The way to identify whether a couple of bands belong to $J_z=\pm1/2$ and $J_z=\pm3/2$ fermions is by looking at how the states transform under rotational symmetry. By analyzing Supplementary Table~\ref{Table_S1}, we find for $J_z=\pm1/2$ fermions, there are KNLs within the mirror plane or along the roto-inversion axis of S3 and S4 symmetry, which is consistent with the general analysis given in Sec. II; while for $J_z=\pm3/2$ fermions, KNLs are only enforced within the mirror plane. In Sec. V, we further show KNLs of $J_z=\pm3/2$ fermions in a real material as an example.

In addition, it is also interesting to study the dispersion relation between the couple of bands around certain KNLs. By checking Supplementary Table~\ref{Table_S1}, we find that besides linear touching KNLs, there are also quadratic KNLs in the $C_{3h}$ and $D_{3h}$ point groups and cubic KNLs in the $C_{6v}$ point group. 

Here the terms: linear, quadraticand cubic are defined by the dispersion of splitting between two bands upon a $k$-plane perpendicular to the KNLs that are studied. For a plane in the 3D Brillouin zone which intercepts a KNL, we can obtain a $\mathbf{k\cdot p}$ Hamiltonian describing the states near the KNL on the momentum plane. For example, assuming a KNL along the $k_z$-direction, the $\mathbf{k}\cdot\mathbf{p}$ Hamiltonian near the KNL can be written as
$H(\mathbf{p})=f_0(\mathbf{p})\sigma_0+v\mathbf{p}_+^m\sigma_++v\mathbf{p}_-^m\sigma_-$,	
where $\mathbf{p}$ denotes the momentum perpendicular to the KNL. Here, $\mathbf{p}_\pm=p_x+ip_y, \sigma_\pm=\sigma_x\pm\sigma_y$ and $m$=1,2,3 determines the energy dispersion of the states which gives  linear-, quadratic- and cubic- energy dispersion respectively. We use the word \textit{Dirac} to denote that case with $m=1$ and the word \textit{higher-order Dirac} to denote that cases with $m>1$.  Note that the Dirac point here is the interception point between the momentum plane and the KNL and it is always two-fold degenerate. By adiabatically moving an electron in a loop circling a given KNL, it will acquire a Berry phase of $m\pi$ mod $2\pi$ which can be experimentally probed by quantum oscillation.


\section{Determine Kramers nodal lines from compatibility relations}

In general when moving from a high symmetry point to a high symmetry line, the symmetry of $k$-points is reduced. For convenience, we denote the little group of the high symmetry point as $\mathcal{G}_1$, and its subgroup $\mathcal{G}_2$ as the little group of the high symmetry line. Then an irreducible representation $\Gamma_1$ of $\mathcal{G}_1$ can be decomposed as linear combinations of irreducible representations $\Gamma_j$ of $\mathcal{G}_2$, i.e., the character $\chi$ of each unitary symmetry operation $R$ satisfies 
\begin{equation}
\chi(D^{(\Gamma_1)}_{\mathcal{G}_1}(R))=\sum_j \chi(D^{(\Gamma_j)}_{\mathcal{G}_2}(R))\label{Eq_compatibility},
\end{equation}
This formula defines the compatibility relations. Here we illustrate how to determine Kramers nodal lines via compatibility relations with SG No.~156 ($P3m1$) as an example.  The same method has been used to determine the nodal points and nodal lines enforced by non-symmorphic symmetries in Ref.~\cite{S_Bradlyn2,S_Zhang, S_Chan}.

For SG No.~156 ($P3m1$), relevant symmetries are the three-fold symmetry $3_{001}$ and the mirror symmetry $m_{010}$ ($k_x k_z$-plane).   The mirror plane contains TRIMs $\Gamma$, M, A, L and high symmetry lines $\Delta$, U, R, $\Sigma$. Since the spin-orbital coupling is included, we need to consider double-valued representations, where a $2\pi$ rotation will yield a $-1$ phase. These double-valued irreducible representations are listed in Supplementary Table~\ref{table2}. At TRIMs, there is one additional requirement for irreducible representations: time-reversal invariance.  The two-dimensional representations $\overline{\Gamma}_6$, $\overline{A}_6$ are pseudo-real, so they are time-reversal invariant by themselves \cite{Bradley_book}. All one-dimensional representations at TRIMs are complex and need to be paired up to form time-reversal-invariant representations or so-called co-representations \cite{Bradley_book}. With Supplementary Eq.~(\ref{Eq_compatibility}), we are able to determine how these time-reversal invariant representations are split along high symmetry lines. The compatibility relations and the corresponding band connectivity are drawn in Supplementary Fig.~\ref{fig:figs2}. Evidently, only $\overline{\Gamma}_6$--$\overline{\Delta}_6$--$\overline{A}_6$ is able to support the two-fold degenerate KNL. Alternatively, one can  identify this KNL by consulting the program DCOMPREL on Bilbao Crystallographic Server \cite{S_Elcoro}.

\begin{figure*}
	\centering
	\includegraphics[width=0.8\linewidth]{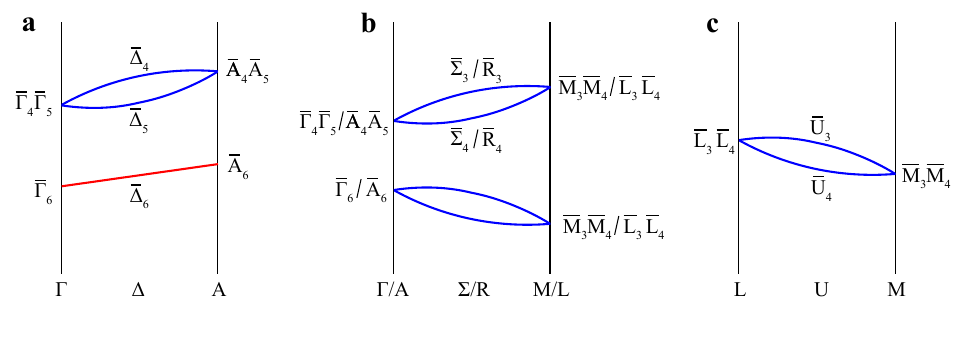}
	\caption{Analysis of KNLs for SG No. 156 ($P3m1$) from compatibility relations. {\bf a}, {\bf b}, {\bf c} demonstrate compatibility relations and band connectivity diagrams for SG No.~156, $P3m1$ along different $k$-path, inferred from Supplementary Table~\ref{table2}. High symmetry lines with two-fold degeneracy are highlighted in red color. }
	\label{fig:figs2}
\end{figure*}

\begin{table}
	\caption{Double-valued irreducible representation (Irrep) of SG No.~156 ($P3m1$) at TRIMs $\Gamma$, A, M, L as well as high symmetry lines $\Delta$, U, R, $\Sigma$. The notations follow from Ref.~\cite{S_Elcoro}.}
	\begin{ruledtabular}\label{table2}
		\begin{tabular}{ccc}
			Irrep & $3_{001}$&$m_{010}$\\\hline
			$\overline{\Gamma}_4$&$-1$& $-i$\\
			$\overline{\Gamma}_5$& $-1$& $i$\\
			$\overline{\Gamma}_6$&$\begin{pmatrix}
			e^{-i\pi/3}&0\\
			0&	e^{i\pi/3}
			\end{pmatrix}$&  $\begin{pmatrix}
			0&e^{-i\pi/3}\\
			e^{-i2\pi/3}&0
			\end{pmatrix}$\\\hline
			$\overline{\Delta}_4$&$-1$&$-i$\\
			$\overline{\Delta}_5$&$-1$&$i$\\
			$\overline{\Delta}_6$& $\begin{pmatrix}
			e^{-i\pi/3}&0\\
			0&e^{i\pi/3}
			\end{pmatrix}$& $\begin{pmatrix}
			0&e^{-i\pi/3}\\
			e^{-i2\pi/3}&0
			\end{pmatrix}$\\\hline
			$\overline{A}_4$&$-1$& $-i$\\
			$\overline{A}_5$&$-1$& $i$\\
			$\overline{A}_6$& $\begin{pmatrix}
			e^{-i\pi/3}&0 \\
			0&	e^{i\pi/3}
			\end{pmatrix}$& $\begin{pmatrix}
			0& e^{-i\pi/3}\\
			e^{-i2\pi/3}&0
			\end{pmatrix}$\\\hline
			$\overline{L}_3$&--&$-i$\\
			$\overline{L}_4$&--&$i$\\\hline
			$\overline{M}_3$&--&$-i$\\
			$\overline{M}_4$&--&$i$\\\hline
			$\overline{U}_3$&--&$-i$\\
			$\overline{U}_4$&--&$i$\\\hline
			$\overline{\Sigma}_3$&--&$-i$\\
			$\overline{\Sigma}_4$&--&$i$\\\hline
			$\overline{R}_3$&--&$-i$\\
			$\overline{R}_4$&--&$i$
		\end{tabular}
	\end{ruledtabular}
\end{table}

Here, we need to comment on a special case where degeneracy cannot be captured by the analysis of ordinary compatibility relations. When the high symmetry line considered is along a roto-inversion axis of $S_n$ ($n=4,6$), such as in SG No.~174 ($P\bar{6}$) and SG No.~81 ($P\bar{4}$), a combined anti-unitary symmetry $\mathcal{T}S_n$ can also enforce degeneracy.  Take SG No.~81 ($P\bar{4}$) as an example. The combined anti-unitary symmetry $\mathcal{T}S_4$ leaves the $k$-points upon $\Gamma$--Z (or equivalently high symmetry line $\Lambda$) invariant. Due to this anti-unitary symmetry, the double-valued complex irreducible representations  $\overline{\Lambda}_3$ and $\overline{\Lambda}_4$  pair up and form a two-dimensional irreducible co-representation, which yields the $\Gamma$--Z KNL in SG No.~81 ($P\bar{4}$). In this case, we can also understand this KNL from the eigenvalue method \cite{S_Bradlyn2}. The Hamiltonian $H$ along $\Gamma$--Z in SG No.~81 ($P\bar{4}$) is actually not only invariant under  the $\mathcal{T}S_4$ operation, but also under the $C_2$ operation. Let us consider a simultaneous eigenstate of $H$ and $C_2$ as $\psi$, with $C_2\psi=\lambda\psi$ and $H\psi=E\psi$. It is easy to show $\mathcal{T}S_4\psi$ is also an eigenstate of $C_2$ with eigenvalue $\lambda^*$ as well as an eigenstate of $H$ with eigenvalue $E$, because we have $[\mathcal{T}S_4$ and $C_2]=0$ and $[\mathcal{T}S_4, H]=0$. The 1/2 spin of electrons further requires $\lambda^2=-1$, leading to $\lambda=\pm i$ and $\lambda=-\lambda^*$. This means $\psi$ and $\mathcal{T}S_4\psi$ are two distinct states with the same eigen-energy $E$ and they form a two-dimensional irreducible co-representation.

The derivation of this special case for other space groups listed in main text Table 1 proceeds in a similar way. All of the allowed KNLs by the space group symmetries are summarized in main text Table 1. They are compatible with the $\mathbf{k\cdot p}$ analysis given in Supplementary Table~\ref{Table_S1}.

\begin{table*}
	\caption[]{The little groups of TRIMs in non-centrosymmetric achiral symmorphic space groups.	}
	\begin{ruledtabular}\label{Table_S3}
		\begin{tabular}{ccc}
			SG No. & TRIMs with  achiral little group& TRIMs with chiral little group\\
			\hline
			6,$Pm$ & $\Gamma$, $B$, $Y$, $A$, $Z$, $C$, $D$ and $E$ are all $C_{1h}$.&--\\
			8,$Cm$ & $\Gamma$, $Y$, $A$ and $M$ are all  $C_{1h}$.&--\\
			25,$Pmm2$ &$\Gamma$, $Z$, $Y$, $T$, $X$, $U$, $S$ and $R$ are all $C_{2v}$.&--\\
			35,$Cmm2$ & $\Gamma$, $Z$, $Y$ and $T$ are all $C_{2v}$.& $S$ and $R$ are both $C_{2}$.\\
			38,$Amm2$ & $\Gamma$, $Y$, $T$, $Z$ and are all $C_{2v}$.&--\\
			42,$Fmm2$ &$\Gamma$, $Z$, $T$ and $Y$ are all $C_{2v}$.&--\\
			44,$Imm2$ & $\Gamma$ and $X$ are $C_{2v}$; $S$ and $R$ are $C_{1v}$.&$T$ is $C_{2}$.\\
			81,$P\bar{4}$ & $\Gamma$, $Z$, $M$ and $A$ are all $S_4$& $X$ and  $R$  are both $C_2$.\\
			82,$I\bar{4}$ &$\Gamma$, $M$ are both $S_4$.&$N$ is $C_1$, $X$ is $C_2$.\\
			99,$P4mm$ &$\Gamma$, $Z$, $M$ and $A$ all are  $C_{4v}$;$X$ and $R$ are both  $C_{2v}$.&--\\
			107,$I4mm$ & $\Gamma$ and $M$ are both $C_{4v}$; $X$ is $C_{2v}$; N is $C_{1v}$.&--\\
			111,$P\bar{4}2m$ & $\Gamma$, $A$, $Z$ and $M$ are all $D_{2d}$.& $X$ and $R$ are both $D_2$.\\
			115,$P\bar{4}m2$ & $\Gamma$, $M$, $A$ and $Z$  are all $D_{2d}$; $X$ and $R$ are both $C_{2v}$.&--\\
			119,$I\bar{4}m2$ & $\Gamma$ and $M$ are both $D_{2d}$, $N$ is $C_{1v}$. & $X$ is $D_2$.\\
			121,$I\bar{4}2m$ & $\Gamma$ and $M$ are both $D_{2d}$, $X$ is $C_{2v}$.&$N$ is $C_2$.\\
			156,$P3m1$ &$\Gamma$ and $A$ are both $C_{3v}$; $M$ and $L$ are $C_{1v}$.&--\\
			157,$P31m$ & $\Gamma$ and $A$ are both $C_{3v}$; $M$ and $L$ are both $C_{1v}$.&--\\
			160,$R3m$ & $\Gamma$ and $T$ are both $C_{3v}$; $L$ and $FA$ are both $C_{1v}$.&--\\
			174,$P\bar{6}$ & $\Gamma$ and $A$ are both $C_{3h}$; $M$ and $L$ are both $C_{1v}$.&--\\
			183,$P6mm$ & $\Gamma$ and $A$ are both $C_{6v}$; $M$ and $L$ are both $C_{2v}$.&--\\
			187,$P\bar{6}m2$ & $\Gamma$ and $A$ are both $D_{3h}$; $M$ and $L$ are both $C_{2v}$.&--\\
			189,$P\bar{6}2m$ & $\Gamma$ and $A$ are both $D_{3h}$; $M$ and $L$ are both $C_{2v}$.&--\\
			215,$P\bar{4}3m$ & $\Gamma$ and $R$ are both $T_d$; $M$ and $X$ are both $D_{2d}$.&--\\
			216,$F\bar{4}3m$ & $\Gamma$ is $T_d$; $X$ is $D_{2d}$; $L$ is $C_{3v}$.&--\\
			217,$I\bar{4}3m$ & $\Gamma$ and $H$ are both $T_d$.&--
		\end{tabular}
	\end{ruledtabular}
\end{table*}
\section{TRIMs in achiral crystals}
\subsection{An overview of symmorphic space groups}

There are in total 73 symmorphic space groups. Among these 73 space groups, there are 21 centrosymmetric  space space groups: $C_i$:  2; $C_{2h}$: 10 and 12; $D_{2h}$:  47, 65, 69 and 71; $C_{4h}$: 83 and 87;  $D_{4h}$: 123 and  139; $C_{3i}$: 147 and 148; $C_{6h}$: 175; $D_{6h}$: 191; $T_h$: 200 and 202, 204; $O_h$:221, 225 and 229;

27 non-centrosymmetric chiral space groups:  
$C_1$:  1; $C_2$: 3 and 5; $D_2$: 16, 21, 22 and 23; $C_4$: 75 and 79; $D_4$: 89 and 97; $C_3$: 143 and 146;  $D_3$:  149, 150 and 155; $D_{3d}$: 162, 164 and 166; $C_6$: 168; $D_6$: 177; $T$
: 195, 196 and 197; $O$: 207, 209 and 211;

and 25 non-centrosymmetric achiral space groups: $C_{1h}$: 6 and  8; $C_{2v}$: 25, 35, 38, 42 and 44; $S_4$: 81 and 82; $C_{4v}$: 99 and 107;   $D_{2d}$: 111, 115, 119 and 121; $C_{3v}$: 156, 157 and 160;
$C_{3h}$: 174; $C_{6v}$: 183; $D_{3h}$: 187 and 189; $T_d$: 215, 216 and 217.
\subsection{Little groups of TRIMs in achiral space groups}
In the main text, Type I and Type II KNLMs are identified according to the little groups of TRIMs. Here, we summarize the little groups of all TRIMs in  non-centrosymmetric achiral symmorphic space groups in Supplementary Table~\ref{Table_S3}.  The little group of each TRIM can be identified by consulting the program KVEC and MKVEC on the Bilbao Crystallographic Server \cite{S_Elcoro, S_YXu, S_Elcoro_arxiv}.  Apparently, among the 25 non-centrosymmetric symmorphic achiral space groups, seven space groups (SG No.~35, 44, 81, 82, 111, 119 and 121) support chiral TRIMs, while the other eighteen ones do not. As discussed in the main text, these chiral TRIMs that host electronic states described by a two-dimensional irreducible corepresentations of their little groups will emerge as Kramers Weyl points in achiral crystals.  Notably, the appearance of Kramers Weyl points at high-symmetry points in achiral crystals was introduced in Ref.~\cite{S_Chang} as well.

\section{Model Hamiltonians for Rashba semiconductor BiTeI and The circular photogalvanic effect in strained BiTeI}

\subsection{A four-band effective low energy model for BiTeI}

Here, we derive the model Hamiltonians of BiTeI that are used in the main text. The lattice structure of BiTeI belongs to SG No.~156 ($P3m1$), the point group of which is a polar point group $C_{3v}$ which is generated by a three-fold rotation $C_3$ along the $z$ axis (\textit{i.e.} $3_{001}$) 
and vertical mirror symmetry $\sigma_v$ (\textit{i.e.} $m_{010}$). 
According to the \textit{ab initio} method, the bands lying closest to Fermi energy are $\ket{\Lambda, p_z, J_z=\pm 1/2}$ bands, where $\Lambda$=Bi,Te,I 
\cite{S_Bahramy}. By analyzing the transformation properties of $\ket{\Lambda,p_z,\pm 1/2}$, the four bands near Fermi energy at $A$ point are found to belong to the spinor irreducible representation $\overline{\Gamma}_6$ of double group $C_{3v}$.

\begin{table}
	\caption{Model parameters for effective Hamiltonian \ref{four_Ha2}. The chemical potential  $\mu$ is chosen to  be near the crossing point of the conduction band. }
	\begin{tabular}{ccccc}
		\hline\hline
		$C_0^{c}$(eV)&$C_1^{c}$(eV$\cdot$\AA$^2$)&$C_2^{c}$(eV$\cdot$\AA$^2$)&$\alpha_0^{c}($eV$\cdot$\AA)&$\alpha_1^{c}($eV$\cdot$\AA$^3$)\\
		\hline
		0.2491 &24.0587&3.2389&1.4687&-18.2993\\
		\hline\hline
		$C_0^{v}$(eV)&$C_1^{v}$(eV$\cdot$\AA$^2$)&$C_2^{v}$(eV$\cdot$\AA$^2$)&$\alpha_0^{v}($eV$\cdot$\AA)&$\alpha_1^{v}($eV$\cdot$\AA$^3$)\\
		\hline
		0.0757&-4.4977&-7.7134&-0.0982&-0.5719\\
		\hline\hline
		$\mathcal{M}_0^0$ (eV)&$\mathcal{M}_1^0$(eV$\cdot$\AA)&$\mathcal{M}_2^0$(eV$\cdot$\AA$^2$)&$\mathcal{A}^0$(eV$\cdot$\AA)&$\mathcal{B}^0$(eV$\cdot$\AA)\\\hline
		0.2362&-6.6320&2.5584&0.3689&2.3023\\
		\hline\hline
		$\mathcal{D}^0$(eV$\cdot$\AA$^2$)&$\mu$(eV)&$a$ (\AA)&$c$ (\AA)&\\
		\hline
		1.4725&0.1151&4.425&7.378&\\\hline
	\end{tabular}	
	\label{Model_para}
\end{table}

Based on this symmetry analysis, we can construct a four-band low energy effective Hamiltonian:
\begin{equation}
H_{eff}(\mathbf{k})=\begin{pmatrix}
\epsilon_1(\mathbf{k})&V_0(\mathbf{k})\\
V_0^{\dagger}(\mathbf{k})&\epsilon_2(\mathbf{k})
\end{pmatrix}. \label{four_Ha2}
\end{equation}
Here 
\begin{equation}
\begin{split}
\epsilon_i(\mathbf{k})&=C_0^i+C_1^ik_{\parallel}^2 +C_2^ik_z^2\\
&+(\alpha_0^i+\alpha^i_1k^2_{\parallel})(k_x\sigma_y-k_y\sigma_x)
+\beta^{i}k_y(3k_x^2-k_y^2)\sigma_z
\end{split}
\end{equation} 

\begin{equation}
\begin{split}
V_0(\mathbf{k})&=\mathcal{M}_0+\mathcal{M}_1k_{\parallel}^2+\mathcal{M}_2k_z^2-i\mathcal{A}k_z\\
&+\mathcal{B}(k_x\sigma_y-k_y\sigma_x)-i\mathcal{D}((k_x^2-k_y^2)\sigma_y+2k_xk_y\sigma_x)
\end{split}
\end{equation} 
with $k_{\parallel}^2=k_x^2+k_y^2$ . The values of these parameters can be determined by fitting the unstrained BiTeI DFT band structure, which are listed in Supplementary Table~\ref{Model_para}. 

\begin{figure*}
	\centering
	\includegraphics[width=1\linewidth]{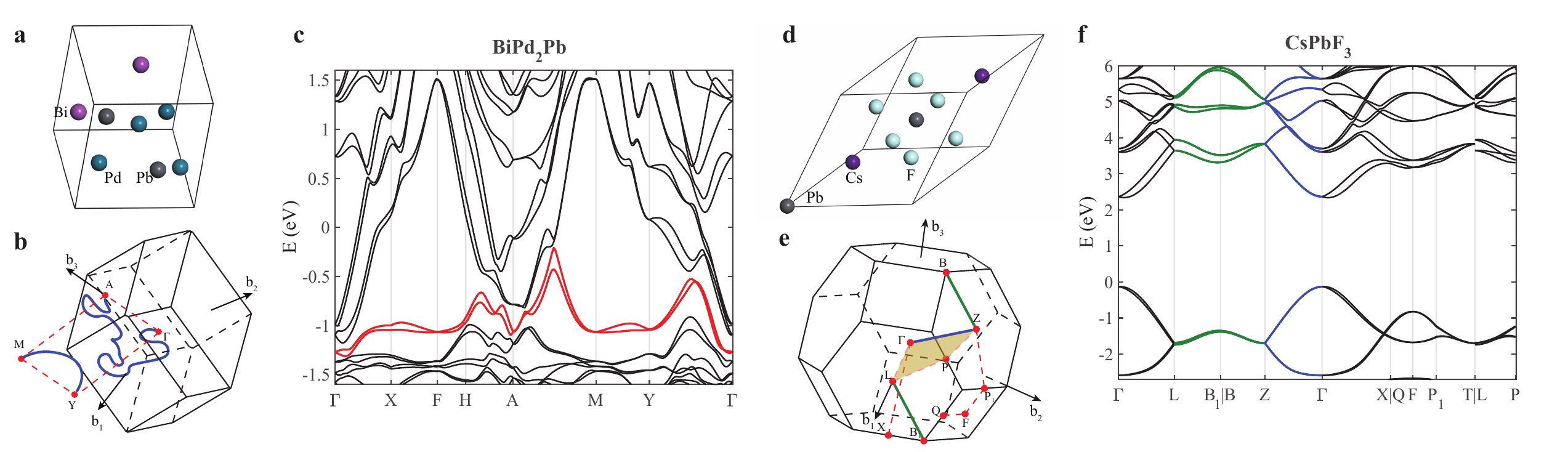}
	\caption{Example materials of crystals with the $C_{1v}$ and nonsymmorphic symmetry.
		{\bf a} and {\bf b} show the primitive cell and the first Brillouin zone of BiPd$_2$Pb, respectively. The DFT calculated BiPd$_2$Pb band structures is presented in {\bf c}. The degenerate KNLs given by the two red-colored bands in {\bf c} are plotted as blue curves in {\bf b}.
		{\bf d, e} and {\bf f} respectively show the non-symmorphic material CsPbF$_3$\rq{}s primitive lattice cell, first Brillouin zone and DFT bands, respectively. The thick blue and green curves in {\bf e} and {\bf f} are the KNLs connecting $\Gamma$-T and T-L, where KNL L-B-T enforced by the glide mirror symmetry is highlighted as green color.}
	\label{fig:figs3}
\end{figure*}	
\subsection{The circular photogalvanic effect  of strained BiTeI}

The circular photogalvanic effect (CPGE) describes the DC part of the photocurrent produced by circularly polarized light which reverses sign when circular polarization is reversed. The quantization of CPGE is a signal for the emergence of Kramers Weyl points \cite{S_Chang}. 
The the chiral charge $\mathcal{C}$  shown in main text Fig.4 is defined as \cite{S_Sipe,S_Moore}
\begin{equation}
\mathcal{C}=\text{Tr}(\beta)/i\beta_0
\end{equation}
with  CPGE tensor 
\begin{equation}
\beta_{ij}(\omega)=\frac{\pi e^3}{\hbar V}\epsilon_{ijk}\sum_{\mathbf{k},n,m} f_{nm}^{\mathbf{k}}\Delta_{\mathbf{k},nm}^ir_{\mathbf{k},nm}^{k}r_{\mathbf{k},mn}^{l}
\delta(\hbar\omega-E_{\mathbf{k},mn}).
\end{equation}
Here, $\beta_0=\pi e^3/h^2$, $V$ is the sample volume, $r_{\mathbf{k},nm}=i\braket{n|\partial_{\mathbf{k}}|m}$  is the Berry connection  between the $n$th and $m$th bands, $E_{\mathbf{k},nm}=E_{\mathbf{k},n}-E_{\mathbf{k},m}$, $f_{nm}^{\mathbf{k}}=f_n^{\mathbf{k}}-f_m^{\mathbf{k}}$ represent the energy difference and Fermi-Dirac distribution, respectively, and $\Delta_{\mathbf{k},nm}^{i}=\partial_{\mathbf{k}_i}E_{\mathbf{k},mn}/\hbar$ is the electron velocity. From this formula,  we calculated the trace of CPGE tensor of a strained BiTeI, which is captured by the Hamiltonian $H_{eff}(\mathbf{k})+H_{strain}$. The strained effects are described by a mirror-broken phenomenological Hamiltonian $H_{strain}=\lambda k_z \sigma_z$. We estimate the value of $\lambda$ by fitting the splitting of $\Gamma$--Z  in  a strained band structure with $\lambda=5$ meV,  $15$meV and $25$ meV for $1\%$, $3\%$ and $5\%$ strains, respectively.  The Fig.~4 in the main text was calculated using this model, and the chemical potential has been set near the Weyl nodes of the conduction band. Note that the influence of valence bands on the CPGE has also been taken into consideration in this four-band model, while we found that  this influence is actually negligible within the low frequency region, which is consistent with the result in Ref.~\cite{S_Moore}.

\begin{figure*}
	\centering
	\includegraphics[width=1\linewidth]{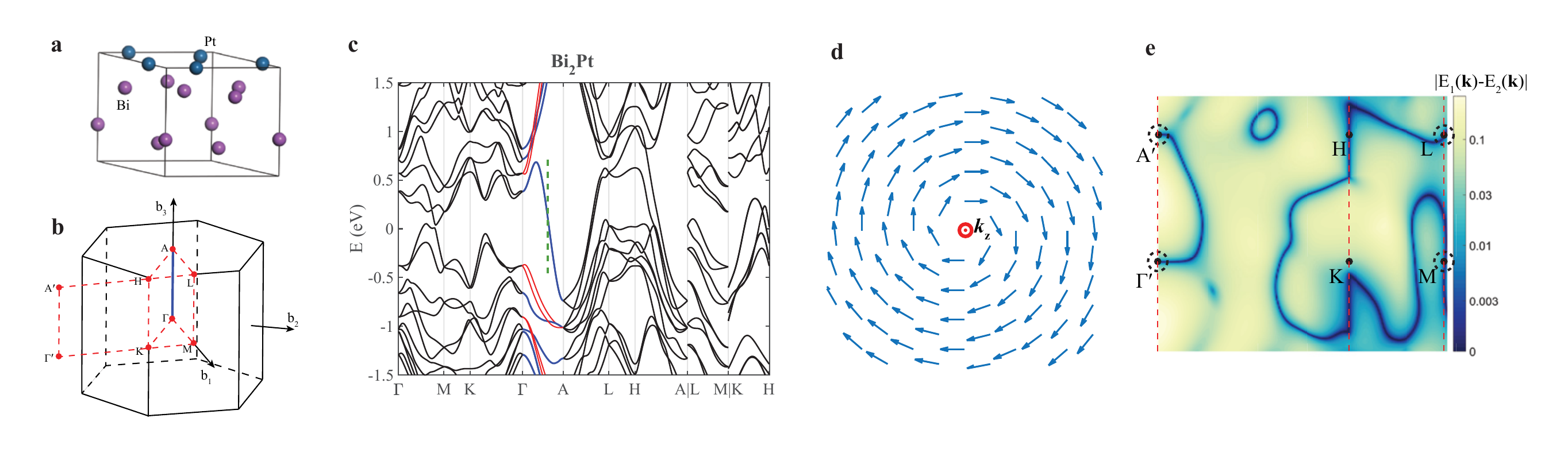}
	\caption{Bi$_2$Pt (SG No.~157, $P31m$): a KNLM with $J_z=\pm3/2$ bands.  {\bf a, b} and {\bf c}  show Bi$_2$Pt\rq{}s primitive lattice cell, 1st Brillouin zone and its DFT bands, respectively. The thick blue and red curves along the $\Gamma$-A line in {\bf c} represent $J_z=\pm1/2$ and $J_z=\pm3/2$ bands, respectively. {\bf d} plots the DFT-obtained spin texture on a plane (denoted by the green dashed line in {\bf c}) perpendicular to the KNL $\Gamma$--A. {\bf e} plots the band gap of two Bi$_2$Pt\rq{}s $J_z=\pm3/2$ bands on the mirror-invariant plane upon which contains four TRIMs, $\Gamma^\prime$, A$^\prime$, L and M.  Note that H and K are two high symmetry points but not TRIMs on the mirror plane. The mirror plane is illustrated in {\bf b}. Two KNLs connecting $\Gamma'$--A$^\prime$ and M--L can be easily identified.}
	\label{fig:figs4}
\end{figure*}

\begin{figure*}
	\centering
	\includegraphics[width=1\linewidth]{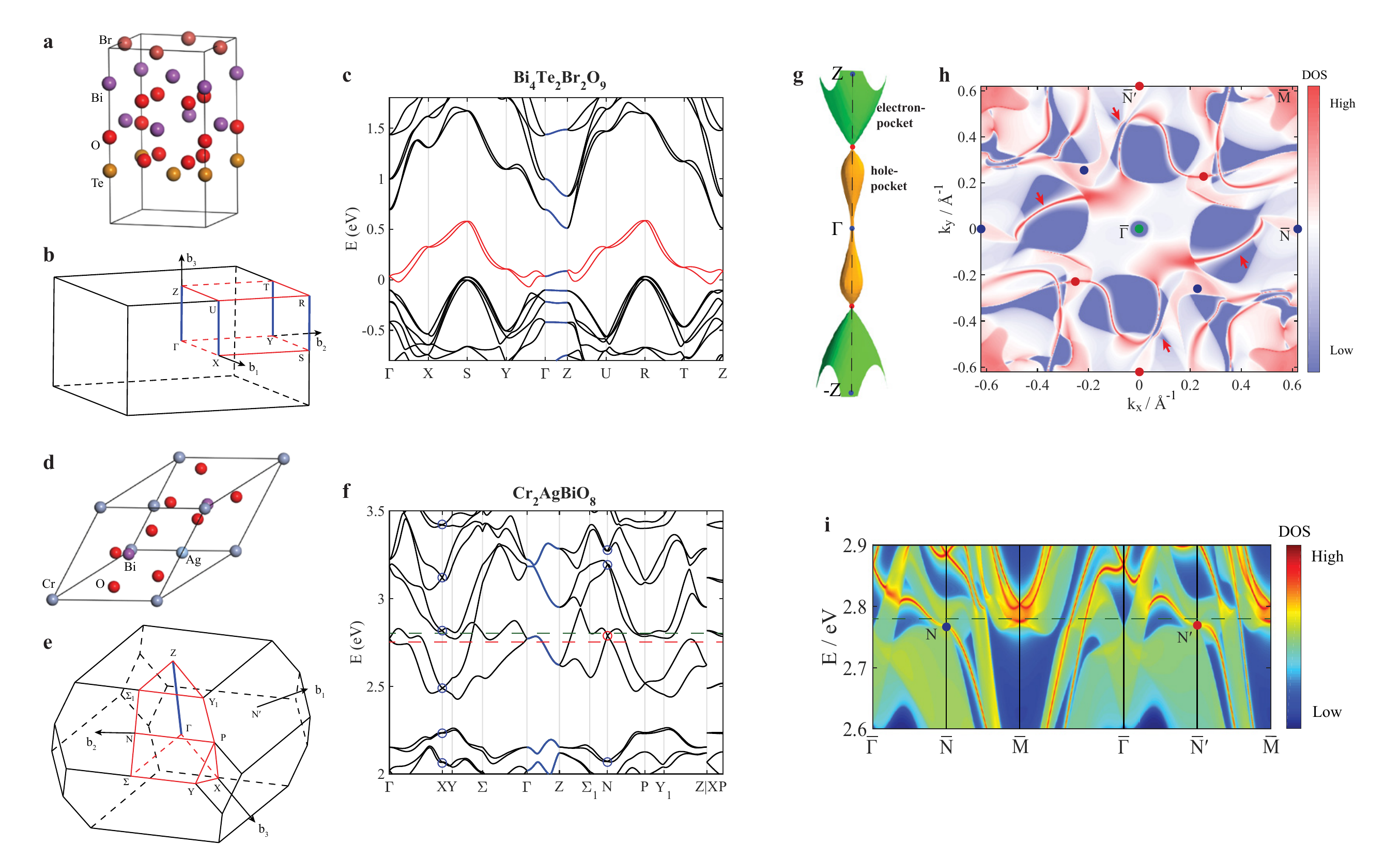}
	\caption{ Example materials of type I and type II KNLM with the octdong Fermi surface.
		In {\bf a, b} and {\bf c} we show the primitive lattice cell, the first Brillouin zone and the conduction bands, respectively, of the KNLM,  Bi$_4$Te$_2$Br$_2$O$_9$, which possesses octdong Fermi surfaces as mentioned in the main text Sec. II C. The KNLs along $\Gamma$--Z are denoted by thick blue lines in both {\bf b} and {\bf c}. 
		In {\bf d, e} and {\bf f} we illustrate the primitive lattice cell, the first Brillouin zone and the conduction bands of the Type II KNLM Cr$_2$AgBiO$_8$.  As a Type II KNLM, Cr$_2$AgBiO$_8$ has Kramers Weyl points N (N$^\prime$) and X, which are circled in {\bf f}. In {\bf g}, the octdong-type Fermi surface resulting from the $\Gamma$--Z KNL are drawn at $E_F=2.768$eV, as denoted by the red dashed line in {\bf f} (some trivial Fermi surfaces near this octdong Fermi surface are not depicted in {\bf g}). Here the red dots denote the touching points of the electron pocket centered at Z and the hole pocket centered at $\Gamma$. In {\bf h}, we plot the (001) surface spectral function of Cr$_2$AgBiO$_8$ at $E=2.78$eV (denoted as a green dashed line in both {\bf f} and1 {\bf i}). The blue (red) solid dots represent the Weyl points with negative (positive) chiral charges. Note that the Kramers Weyl point N (N$^\prime$) projects to $\bar{\mathrm{N}}$ ($\bar{\mathrm{N}}^\prime$) on the surface Brillouin zone. The four Fermi arcs originating from the N and N$^\prime$ pockets are pointed out by red arrows. {\bf i} is the surface spectral function along a k-path on the surface Brillouin zone.}
	\label{fig:figs5}
\end{figure*}

\section{More representative materials of KNLMs}

In this section, we list some representative materials of KNLMs, including some special cases which have not been discussed in the main text.

\subsection{BiPd$_2$Pb: $C_{1v}$ system with only one mirror}

The band structure of BiPd$_2$Pb (SG No.~8, $Cm$, point group $C_{1v}$) is shown in Supplementary Fig.~\ref{fig:figs3}{\bf c}. The two bands depicted in red are used to plot Fig.~2{\bf j} of the main text. Unlike the cases of most of the material listed in main text Table 1, where all KNLs are pinned along high-symmetry paths by the crystal symmetry, 
the single mirror symmetry in $C_{1v}$ only restricts the KNLs to lie within the mirror-invariant $k$-plane but not necessarily along high-symmetry paths, as shown in Supplementary Fig.~\ref{fig:figs3}{\bf b} and Supplementary Fig.~\ref{fig:figs3}{\bf c}. A more clear figure of the KNLs which is directly related to the DFT results has been presented in the main text Fig.~2.

\subsection{CsPbF$_3$: A Nonsymmorphic KNLM}

In the main text, we have constrained our discussion within the symmorphic crystals as some non-symmorphic symmetry could give rise to nodal planes on the Brillouin zone boundaries. In the following, a non-symmorphic KNLM CsPbF$_3$ is given as an example to illustrate that our discussion can also be applied to non-symmorphic crystals. CsPbF$_3$ belongs to the non-symmorphic SG No.~161 which is also denoted as $\Gamma_{rh}C_{3v}^6$ or $R3c$. As shown in Supplementary Fig.~\ref{fig:figs3}{\bf e}, and Supplementary Fig.~\ref{fig:figs3}{\bf f}, there are KNLs along the $\Gamma$--Z path in CsPbF$_3$, which is similar to the case of symmorphic SG No.~160 ($R3m$) shown in the main text Table 1. In addition, there is also a KNL connecting TRIMs Z and L via point $B/B_1$, which is denoted by the green lines in the corresponding figures. It should be noted that due to the lack of screw symmetries in SG No.~161 ($R3c$), there is no non-symmorphic symmetry forcing nodal planes on the Brillouin zone boundary for the case of CsPbF$_3$, which was verified by DFT calculations.   As discussed in Ref.~\cite{S_Chang}, nodal degeneracies at the zone-boundary $k_i=\pi/a_i$ can be supported if the screw symmetry $\{C_{2,i}|\mathbf{t}\}$ with $\mathbf{t}_i=a_i/2$ ($a_i$ as lattice constant along $i$ axis) being contained in the SG symmetry $G$. This is because a combined anti-unitary symmetry $\{C_{2,i}|\mathbf{t}\}\mathcal{T}$ can be defined at the Brillouin zone boundary $k_i=\pi/a_i$, which leaves $\mathbf{k}$ to be invariant at this plane and 
\begin{eqnarray}
(\{C_{2,i}|\mathbf{t}\}\mathcal{T})^2&&=e^{-i(\frac{\hat{k}_ia_i}{2}+\frac{\hat{k}_ja_j}{2})}C_{2,i}\mathcal{T}e^{-i(\frac{\hat{k}_ia_i}{2}+\frac{\hat{k}_ja_j}{2})}C_{2,i}\mathcal{T}\nonumber\\
&&=e^{-ik_ia_i}=-1.
\end{eqnarray}
Here, the index $j$ labels the other two components that are orthogonal to $k_i$,  the crystal momentum operator respects $\mathcal{T}\hat{k}_i \mathcal{T}^{-1}=-\hat{k}_i$. These additional nodal degeneracies at $k_i=\pi/a_i$ overwhelm the KNLs on this plane.

Although the glide symmetry is not related to the nodal plane degeneracies at Brillouin zone boundaries, we found it can enforce some extra KNLs at Brillouin zone boundaries. Especially,  the direction of the KNL is perpendicular to the glide mirror plane in this case. To be specific, we consider a glide mirror symmetry $\{m_1|\mathbf{t}\}$, where the translational operation  $\mathbf{t}=(a_{1}/2,a_{2}/2, a_{3}/2)$ with $a_{2,3}/2 (a_{1}/2)$ as the translations within along (perpendicular to) the mirror plane, and the Brillouin zone boundary is taken as $k_3=\pi/a_3$. Then we show the combined symmetry  $\{m_1|\mathbf{t}\}\mathcal{T}$ is a well-defined anti-unitary symmetry for the line $\mathbf{k}=(k_1, 0, \pi/a_3)$, which lies at the Brillouin zone boundary $k_{3}=\pi/a_3 $ and is perpendicular to the glide mirror plane.  It can be noted that this line is invariant under this combined symmetry operation $\{m_1|\mathbf{t}\}\mathcal{T}$ and on this line, the square of $\{m_1|\mathbf{t}\}\mathcal{T}$ is given by
\begin{equation}
(\{m_1|\mathbf{t}\} \mathcal{T} )^2
=e^{-i(k_2a_2+k_3a_3)}m_1^2\mathcal{T}^2=-1
\end{equation}
Hence, we can see that a glide mirror symmetry enforces a degenerate line that is perpendicular to the glide mirror plane at the Brillouin zone zone boundary for non-magnetic crystals. Indeed, the KNL  connecting TRIMs Z and L via point $B/B_1$  shown in Supplementary Fig.~\ref{fig:figs3}{\bf e}, and Supplementary Fig.~\ref{fig:figs3}{\bf f} is enforced by the glide mirror symmetry that is along $\Gamma$--Z and perpendicular to Z--B direction. Notice that the nodal line Z--B can be extended to the KNL $Z$--L and  part of the KNL $Z$--L are folded back on the Brillouin zone boundary  as $B_1$--L.

\subsection{Bi$_2$Pt: KNLM with $J_z=\pm3/2$ bands}

Under most circumstances, we have assumed that the electronic states transform as $|J=\frac{1}{2}, J_z=\pm\frac{1}{2}\rangle$ states under symmetry operations. However, as is shown in Supplementary Table~\ref{Table_S1}, it is possible for some point groups ($C_{3v}$, $C_{3h}$, $D_{3h}$ and $C_{6v}$) to have double-valued IRs corresponding to $|J=\frac{3}{2}, J_z=\pm\frac{3}{2}\rangle$ states. Although $|J=\frac{3}{2}, J_z=\pm\frac{3}{2}\rangle$ states transform under mirror $m_z$ the same way as $|J=\frac{1}{2}, J_z=\pm\frac{1}{2}\rangle$ states, leading to the same conclusion that there are KNLs in the mirror-invariant plane, they behave quite differently under roto-inversion $S_3=m_zC_3=IC_6$ with $\varphi=\pi/3\cdot 3=\pi$, as defined in Supplementary Eq.~(\ref{Eq:f_pm}) and Supplementary Eq.~(\ref{Eq:f3}). By applying Supplementary Eq.~(\ref{eq:on_axis}), we find that a finite $f_\pm$ is allowed upon the roto-inversion axis, in contrast to the case of $|J=\frac{1}{2}, J_z=\pm\frac{1}{2}\rangle$, where along this axis lies a KNL. 

In Supplementary Fig.~\ref{fig:figs4}{\bf c} we give a vivid example for the above discussion by showing the DFT band structure of Bi$_2$Pt (SG No.~157, $P31m$, point group $C_{3v}$). In the spectrum, the red curves that split along the $\Gamma$-A line represent the $J_z=\pm3/2$ bands, while blue curves, which are doubly-degenerate KNLs along $\Gamma$-A, belong to the $J_z=\pm1/2$ bands. Although the KNLs of $J_z=\pm3/2$ bands are not along the high-symmetry paths, they still exist in the mirror-invariant planes as shown in Supplementary Fig.~\ref{fig:figs4}{\bf e}, which is consistent with the prediction  made in Supplementary Note 3.

\begin{table*}
	\caption{The symmetry-allowed higher-dimensional corepresentations at TRIMs for non-magnetic non-centrosymmetric achiral crystals. }
	\begin{tabular}{c|c|c|c||c|c|c|c}
		\hline\hline
		\multicolumn{8}{c}{Symmorphic SGs}\\\hline
		SG No.&TRIMs& AGs and coreps under $\mathcal{T}$&$d$&SG No. & TRIMs& AGs and coreps under $\mathcal{T}$&$d$\ \\\hline
		215 $F\overline{4}3m$ ($T_d$)& $\Gamma$, R& $G_{48}^{10}$: $R_8$& 4& 216 $F\overline{4}3m$ ($T_d$)&$\Gamma$&$G_{48}^{10}$: $R_8$&4\\\hline
		217 $I\overline{4}3m$ ($T_d$)&$\Gamma$, H&$G_{48}^{10}$: $R_8$&4&&&&\\\hline\hline
		\multicolumn{8}{c}{Nonsymmorphic SGs}\\\hline
		SG No.&TRIMs& AGs and coreps under $\mathcal{T}$&$d$&SG No. & TRIMs& AGs and coreps under $\mathcal{T}$&$d$\\\hline
		26 $Pmc2_1$, 27 $Pcc2$ ($C_{2v}$)& Z,U,T& $G_{16}^8$: $R_9R_9$&4& 29 $Pca2_1$ ($C_{2v}$)& Z, U& $G_{16}^8: R_9R_9$&4\\\hline
		30 $Pnc2$, 31 $Pmn2_1$ ($C_{2v}$)& Z, U& $G_{16}^8: R_9R_9$&4& 32 $Pba2$ ($C_{2v}$) &S, R& $G_{16}^8: R_9R_9$&4\\\hline
		33 $Pna2_1$ ($C_{2v}$) &Z, S, R& $G_{16}^8: R_9R_9$&4&34 $Pnn2$ ($C_{2v}$) &Z, S& $G_{16}^8: R_9R_9$&4\\\hline
		36 $Cmc2_1$ ($C_{2v}$) &Z, T& $G_{16}^8: R_9R_9$&4&37 $Ccc2$ ($C_{2v}$) &Z, T& $G_{16}^8: R_9R_9$&4\\\hline
		43 $Fmm2$ ($C_{2v}$)& Z& $G_{16}^8: R_9R_9$&4& 101 $P4_2cm$ ($C_{4v}$)& R& $G_{16}^8: R_9R_9$& 4\\\hline
		103 $P4cc$ ($C_{4v}$)& Z, A& $G_{32}^{10}: R_6R_6, R_7R_7$& 4& 103 $P4cc$ ($C_{4v}$)& R& $G_{16}^8: R_9R_9$& 4\\\hline
		104 $P4nc$ ($C_{4v}$)& Z&  $G_{32}^{10}: R_6R_6, R_7R_7$& 4& 106 $P4_2bc$ ($C_{4v}$)& A&  $G_{32}^{10}: R_6R_6, R_7R_7$& 4\\\hline
		114 $P\overline{4}2_1c$ ($D_{2d}$)& A& $G_{32}^{10}: R_6R_6, R_7R_7$&4& 116 $P\overline{4}c2$ ($D_{2d}$)& R& $G_{16}^{8}: R_9R_9$& 4\\\hline
		159 $P31c$ ($C_{3v}$)& A& $G_{12}^4: R_5R_5$&4& 161 $R3c$ ($C_{3v}$)& Z& $G^{4}_{12}:R_5R_5$& 4\\\hline
		184 $P6cc$ ($C_{6v}$)& L& $G^{8}_{16}: R_9R_9$& 4&184 $P6cc$ ($C_{6v}$)& A& $G^{12}_{48}: R_7R_7, R_8R_8, R_9R_9$& 4\\\hline
		185 $P6_3cm$, 186 $P6_3mc$ ($C_{6v}$)& L; A& $G^{8}_{16}: R_9R_9$;$G^{13}_{48}: R_{14}R_{15}$& 4& 188 $P\overline{6}c2$ ($D_{3h}$)& A& $G^{14}_{48}: R_{11}R_{12}$&4\\\hline
		190 $P\overline{6}2c$ ($D_{3h}$)& A& $G^{14}_{48}: R_{11}R_{12}$&4& 218 $P\overline{4}3n$ ($T_d$)&$\Gamma$; X& $G^{10}_{48}: R_8$; $G^{11}_{32}: R_6R_7$&4\\\hline
		218 $P\overline{4}3n$ ($T_d$)&R& $G^{7}_{96}: R_6R_7$&4& 219 $F\overline{4}3c$ ($T_d$)&$\Gamma$& $G^{10}_{48}: R_8$&4\\\hline
		220 $I\overline{4}3d$ ($T_d$)&$\Gamma$& $G^{10}_{48}: R_8$&4&220 $I\overline{4}3d$ ($T_d$)&H& $G^{7}_{96}: R_6R_7$&4\\\hline
		218 $P\overline{4}3n$ ($T_d$)& R& $G_{96}^7: R_{15}R_{15}$& 8& 220 $I\overline{4}3d$ ($T_d$)& H& $G_{96}^7: R_{15}R_{15}$& 8\\\hline\hline
	\end{tabular}
	\label{TbS5}
\end{table*}

\subsection{Bi$_4$Te$_2$Br$_2$O$_9$: Type I KNLM with the octdong Fermi surface}

In the main text, we have already mentioned the Type I KNLM Bi$_4$Te$_2$Br$_2$O$_9$ (SG No.~25, $Pmm2$) with four separate KNLs. In Supplementary Fig.~\ref{fig:figs5}{\bf c}, we further show its DFT band structures, within which the couple of bands represented by red curves are the ones related to the octdong Fermi surfaces plotted in the main text Fig.~3.

\begin{figure*}
	\centering
	\includegraphics[width=1\linewidth]{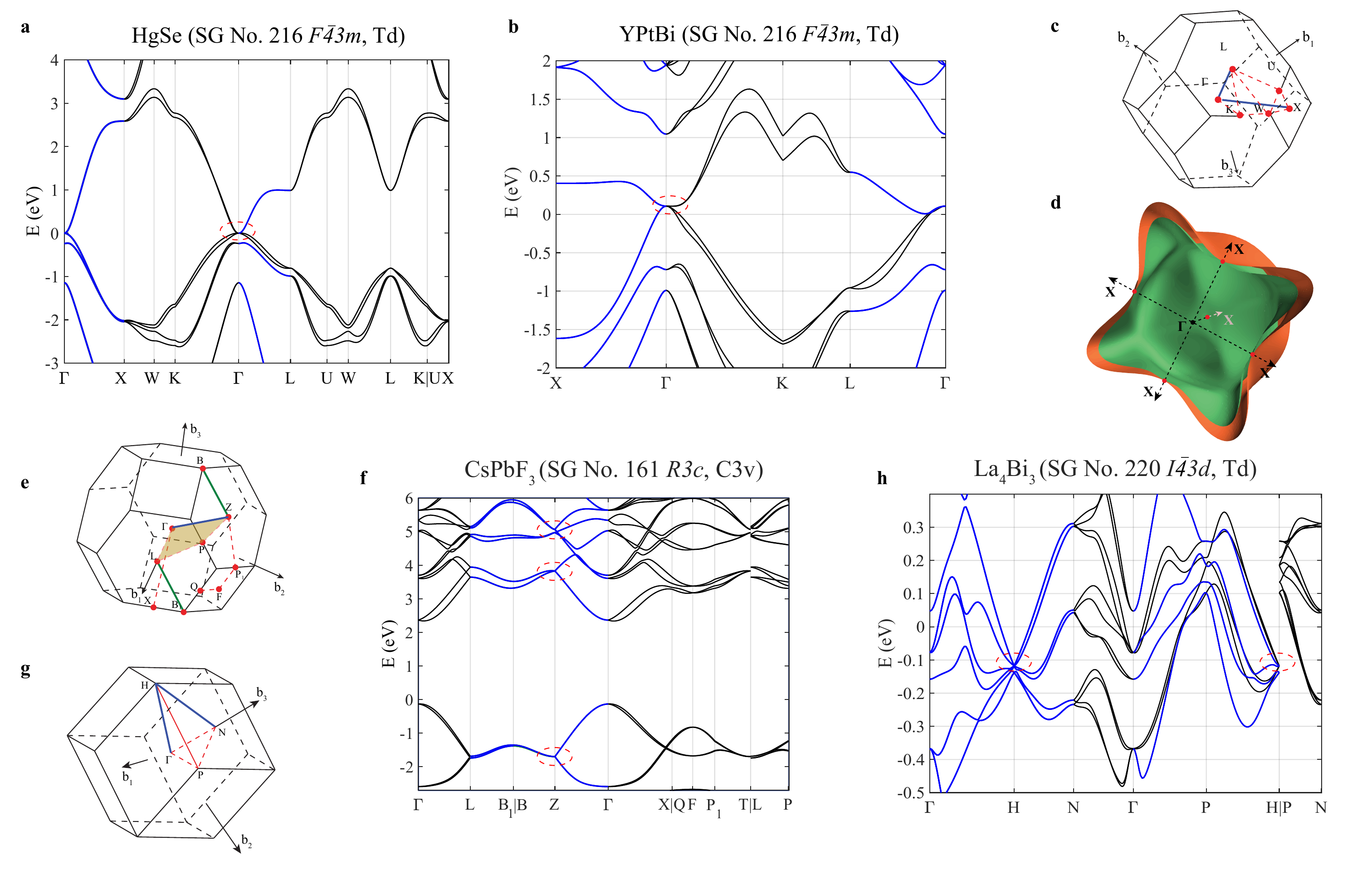}
	\caption{Example materials that exhibit higher copres near Fermi level. {\bf a} and {\bf b}, respectively, show the band structures of symmorphic crystals HgSe (SG No. 216 F$\overline{4}$ 3m,$T_d$), half-Heuser material YPtBi (SG No. 216 F$\overline{4}$3m,$T_d$). {\bf c} and {\bf d}, respectively, show the the 1st Brillouin zone of HgSe, YPtBi and Fermi surface of HgSe at $E=-0.3$eV upon which the red dots denote the touching points. {\bf e}, {\bf g} and {\bf f}, {\bf h}, respectively, show the 1st Brillouin zone and the band structures of nonsymmorphic crystals CsPbF$_3$ (SG No. 161 R3c, $C_{3v}$), La$_4$Bi$_3$ (SG No. 220 I$\overline{4}$3d, $T_d$). The energy bands at TRIMs described by 4D corepresentations are circled in ({\bf a},{\bf b},{\bf f})  and described by 8D corepresentations are circled in {\bf h}. And KNLs are depicted with blue color in  band structures and Brillouin zones}
	\label{fig:figs6}
\end{figure*}

\subsection{Cr$_2$AgBiO$_8$: Type II KNLM with the octdong Fermi surface}

Cr$_2$AgBiO$_8$ (SG No.~82, $I\bar{4}$) is a Type II KNLM as listed in main text Table 1, bearing Kramers Weyl points located at X and N (N$^\prime$). Among all the listed Type II KNLMs, this material specially attracts our interest, because all its bands near the Fermi energy are quite flat compared to its huge SOC splitting, as shown in Supplementary Fig.~\ref{fig:figs5}{\bf f}. This feature provides us with an opportunity to observe the octdong Fermi surface and the Fermi arcs originating from the Kramers Weyl points in this Type II KNLM.

Its KNL is along $\Gamma$--Z, as shown in Supplementary Fig.~\ref{fig:figs5}{\bf e}.  By setting the Fermi energy across a KNL ({\textit i.e.} the red dashed line in Supplementary Fig.~\ref{fig:figs5}{\bf f}), the rare octdong Fermi surface resulting from the $\Gamma$--Z KNL can be seen (Supplementary Fig.~\ref{fig:figs5}{\bf g}).

To show the Fermi arc states, we calculated the surface spectral function at the energy level near $E(\mathbf{k}=\mathrm{N})$ with the surface normal vector parallel to $\Gamma$--Z (Supplementary Fig.~\ref{fig:figs5}{\bf h} and Supplementary Fig.~\ref{fig:figs5}{\bf i}).  Along this projection direction, two distinct N (N$^\prime$) points carrying chiral charge $C=-1$ ($+1$) each, are projected onto the same surface $\mathrm{\bar{N}}$ ($\mathrm{\bar{N}^\prime}$) point. This gives rise to two time-reversal related Fermi arcs coming out from the surface $\mathrm{\bar{N}}$ ($\mathrm{\bar{N}^\prime}$) point (as pointed out by the red arrows in Supplementary Fig.~\ref{fig:figs5}{\bf h}). 	Similar to the chiral KWSs, the Fermi arcs in Type II KNLMs are exceptionally long, spanning the entire Brillouin zone as the Kramers Weyl points are well separated in the reciprocal space. Through this example, we demonstrate that the Fermi arcs originating from the Kramers Weyl points are allowed not only in chiral KWSs, but also in achiral Type II KNLMs.

\subsection{The higher-dimensional corepresentations at TRIMs for non-magnetic non-centrosymmetric achiral crystals}
As discussed in main text, there allows higher dimensional corepresentations in some cases. To identify the feature of KNLs in these cases, in this section, we summary symmetry-allowed higher-dimensional corepresentations at TRIMs for non-magnetic non-centrosymmetric achiral crystals (Supplementary Table~\ref{TbS5}) and present some realistic material examples that support higher dimensional corepresentations. 

We explicitly enumerated the possible higher-dimensional corepresentations, which are labeled with the irreducible of abstract groups, allowed by the Herring's little group $^HG^{\mathbf{k}}$ at TRIMs (c.f. \cite{Bradley_book}).    The results are summarized in Supplementary Table~\ref{TbS5}. In symmorphic groups, there allows 4D corepresentations in TRIMs respecting $T_d$ symmetry, including TRIMs $\Gamma$, R in SG No.~215 ($P\bar{4}3m$), $\Gamma$ in SG No.~216 ($F\bar{4}3m$) and $\Gamma$, H in SG No.~217 ($I\bar{4}3m$). In contrast, for  nonsymmorphic achiral SGs, because of the presence of nonsymmorphic operations (glide mirrors or screw rotations) that complicate the algebra, 4D corepresentations are more widely supported at TRIMs. And notably, the TRIM R in SG No.~218 ($P\bar{4}3n$) and the TRIM H in SG No.~220 ($I\bar{4}3d$) further allows 8D corepresentations, which is consistent with the findings of Wieder et al. in \cite{S_Wieder2} and Bradlyn et al. in \cite{S_Bradlyn2}.

Next, we present some realistic material examples to verify the results in Supplementary Table~\ref{TbS5} and show how KNLs emerge out from these TRIMs when higher-dimensional corepresentations are hosted near Fermi level. In Supplementary Fig.~\ref{fig:figs6}, we plotted the band structure of  HgSe (SG No.~216, $F\bar{4}3m$, $T_d$),  half-Heusler material YPtBi (SG No.~216, $F\bar{4}3m$, $T_d$),  CsPbF3 (SG No.~161, $R3c$, $C_{3v}$), La$_4$Bi$_3$ (SG No.~220, $I\bar{4}3d$, $T_d$). The energy bands at TRIMs describing by 4D corepresentations are circled in Supplementary Fig.~(\ref{fig:figs6}{\bf a}, {\bf b}, {\bf f}) and describing by 8D corepresentations are circled in Supplementary Fig.~\ref{fig:figs6}{\bf h}. It can be seen that the appearance of higher corepresentations for these space groups are consistent with Supplementary Table~\ref{TbS5}: The TRIM $\Gamma$ for symmorphic SG No.~216 ($F\bar{4}3m$),  the TRIM Z for nonsymmorphic SG No.~161 ($R3c$) can host 4D corepresentations, while the TRIM H for nonsymmorphic SG No.~220 ($I\bar{4}3d$) can host 8D corepresentations. On the other hand, the KNLs in these band structures are highlighted as blue color. For an achiral noncentrosymmetric space groups, the appearance of high corepresentations at a TRIM can enforce several KNLs touch together at this TRIM (see Supplementary Fig.~\ref{fig:figs6}). It can be found that there always are KNLs emerging out from these achiral TRIMs. And it is worth noting that this is also true for the cases when energy bands on TRIMs are captured by higher-dimensional corepresentations. 


\end{document}